\documentclass[twocolumn]{aa}
\usepackage[referable]{threeparttablex}

\usepackage{amsmath}
\usepackage{newtx}
\usepackage{natbib}
\usepackage{graphicx}
\usepackage{colortbl}
\usepackage{comment}
\usepackage{hyperref}
\hypersetup{
    colorlinks=true,
    linkcolor=black,
    filecolor=black,      
    urlcolor=cyan,
    citecolor=blue,
    }
\renewcommand\makeLineNumber{}

\begin{document}
\title{The Beauty of $k_2$: Probing Stellar Interiors Using Apsidal Motion}
\subtitle{I. The Benchmark Massive Binary HD\,152248}
\authorrunning{Rosu et al.}
\author{S. Rosu\inst{1}\corrauth{sophie.rosu@unige.ch} \and L. Sciarini\inst{1} \email{luca.sciarini@unige.ch} \and S. Ekström\inst{1} \email{sylvia.ekstrom@unige.ch} \and P. Eggenberger\inst{1} \email{patrick.eggenberger@unige.ch}  \and J. Josiek\inst{2} \email{joris.josiek@uni-heidelberg.de}  \and R. Hirschi\inst{3,4} \email{r.hirschi@keele.ac.uk} \and C. Georgy\inst{1} \email{cyril.georgy@unige.ch} }
\institute{Department of Astronomy, University of Geneva, Chemin Pegasi 51, 1290 Versoix, Switzerland  \and Zentrum für Astronomie der Universität Heidelberg, Astronomisches Rechen-Institut, Mönchhofstr. 12-14, 69120 Heidelberg, Germany \and Astrophysics Research Centre, Lennard-Jones Laboratories, Keele University, Keele ST5 5BG, UK \and Kavli IPMU (WPI), The University of Tokyo, 5-1-5 Kashiwanoha, Kashiwa 277-8583, Japan}
\date{}

\abstract{Over the last decades, several independent studies have shown the need for large convective boundary mixing and convective core sizes in massive stars to reproduce a variety of their observed properties. Yet, stars more massive than $20\,M_\odot$ lack a quantitative prescription for convective boundary mixing as well as an unequivocal constraint on the internal mixing mechanisms acting in them.}
{We take advantage that massive stars evolve mainly in binaries to constrain convective boundary mixing.  
We use the apsidal motion observed in main-sequence binary stars -- linked to the internal stellar structure constants $k_2$ of the stars -- to constrain massive stars' internal density stratification and convective core sizes. In this first paper of a series, we develop the methodology and aim at identifying the models that reproduce the stellar parameters of the benchmark twin massive binary HD\,152248.}
{We build stellar evolution models with the GENeva Evolutionary Code (\texttt{GENEC}) assuming two different angular momentum transport schemes: purely hydrodynamic and magneto-diffusive models. We confront single- and binary-star models to assess the impact of tidal locking on the star's evolution. We investigate the impact of convective boundary mixing (overshooting), metallicity, initial helium abundance, initial mass, mass-loss rate, and mixing length parameter on the evolution of stellar parameters.}
{We highlight the $k_2$-discrepancy between models and observations: The $k_2$ from the models are systematically larger than the observed ones. Models predict stars with too low a density contrast between their core and external layers. Both purely hydrodynamic and magneto-diffusive models require large step-overshoot parameters of 1.2-1.3 to reproduce the stellar parameters, including $k_2$. Other parameters have almost no impact. Given the efficiency of tides to synchronise systems, the assumption of pseudo-synchronisation is sound for this system. It sets a constraint on the misalignment angle of stellar rotation axes of $45^\circ-52^\circ$ maximum. Even with such large angles -- unexpected as alignment happens on the same timescale as synchronisation --, it does not solve the $k_2$-discrepancy and a large step-overshoot parameter of at least 1.0 is necessary to reproduce the observations.}
{Reproducing $k_2$ and the apsidal motion rate simultaneously with stellar parameters requires enhanced core boundary mixing. It acts as increasing the main-sequence lifetime of the star and lowering the rate of radius expansion: The star has more time to contrast its core-envelope density profile. Even if the mass-loss rate was underestimated by a factor two, it would have no impact on stellar parameters evolution, including $k_2$. It demonstrates that the apsidal motion is a powerful, robust means to probe stellar interiors. }


\maketitle

\section{Introduction\label{sect:introduction}}
Despite their importance in shaping the Universe, the structure and evolution of massive binaries are still poorly understood. The mechanisms that drive the stars to follow one or another path (merger, supernova type) are key open questions in modern stellar astrophysics, especially in the context of gravitational-wave detections of merging binary black holes or neutron stars. The first gravitational-wave detection in 2015 \citep{abbott16} triggered a boost of activity around the end products of massive binaries and raised issues in our current understanding of mass-loss, binary interactions, and internal mixing.

To correctly predict the fate of massive binaries, it is essential to have a deep, solid understanding of their main-sequence evolution. Most importantly, accurate estimates for the convective core sizes and internal mixing efficiencies are crucial, yet still lacking. Current uncertainties in the predictions of post-main-sequence evolution \citep{martins13}, pre-supernova structure, and core-collapse outcome \citep{ertl16, chieffi20} are primarily caused by the treatment in stellar models of convective boundary mixing \citep{kaiser20} and internal mixing processes responsible for the transport of angular momentum and chemicals in stars. Both are implemented through free, poorly-constrained parameters and assumed constant with stellar mass and age; Yet, their modification significantly affects the main-sequence lifetime and characteristics of the star’s core and compactness, thus final fate. 

Over the last decades, several independent studies have shown the need for larger convective boundary mixing in massive stars in order to: reproduce the main-sequence width in the Hertzsprung-Russell diagram of stellar clusters at high masses \citep{castro14, martinet21, baraffe23}; explain the properties of massive eclipsing binaries with 1D stellar models \citep{tkachenko20, higgins19}, 2\&3D hydrodynamic simulations \citep{baraffe23, mao24}, and 3D non-perturbative modelling \citep{fellay24}; mimic the convective core entrainment in 321D-guided 1D stellar models \citep{scott21}; explain asteroseismic gravity modes of single stars \citep{aerts21, pedersen21}; and avoid the Humphrey-Davidson limit \citep{gilkis21}. \citet{scott21} investigated core entrainment in 1D stellar models and provided a scaling relationship for the step-overshoot parameter $\alpha_\text{ov} = d_\text{ov}/H_\text{P}$ -- where $d_\text{ov}$ is the overshooting distance, that is to say, the extension of the convective core by overshooting, and $H_\text{P}$ is the pressure scale height --, as a function of the initial stellar mass $M_\text{ini}$ combining insight from 3D simulations, 1D stellar models, and spectroscopic observations (their figure 9). For an initial mass of $30\,M_\odot$, they predict $\alpha_\text{ov} = 0.8$. These studies call for an upward revision of the convective boundary mixing in massive stars \citep[values commonly adopted in massive star model grids are in the range $0.1-0.335$, see][]{ekstrom12, limongi06, brott11}.  

In this context, the measure of the apsidal motion in a binary system is a powerful means to probe the interiors of stars. The apsidal motion is the secular precession in time of a close eccentric binary's orbit. For twin systems where the two stars are identical in terms of stellar properties, the apsidal motion rate equation reduces to the expression
\vspace{-0.1cm}
\begin{equation}
\dot\omega=\frac{4\pi k_{2,\star}}{P_\text{orb}} \left(\frac{R_\star}{a}\right)^5 \left[15f+2g\frac{P^2_\text{orb}}{P^2_{\text{rot},\star}}  \right]
 +\left(\frac{2\pi}{P_\text{orb}}\right)^\frac{5}{3} \frac{3(2GM_\star)^\frac{2}{3}}{c^2(1-e^2)},
\label{eqn:omegadot}
\end{equation}
where $G$ and $c$ are the gravitational constant and the speed of light, $f$ and $g$ are functions of the eccentricity given by Eq. (19) of \citet{rosu20a}, $P_{\text{rot},\star}$ is the stellar rotation period, and other parameters are described in Table \ref{table:obs}; the subscript ‘$\star$’ referring to observed stellar properties. The first and second terms in Eq.\,\eqref{eqn:omegadot} correspond to the Newtonian and general relativistic contributions, $\dot\omega_\text{N}$ and $\dot\omega_\text{GR}$, respectively \citep{shakura85}. The former is proportional to the internal stellar structure constant 
\begin{equation}
k_{2,\star} = \frac{3 - \eta_2(R_\star)}{4+2\eta_2(R_\star)},
\end{equation}
where 
\begin{equation}
\eta_2(R_\star) = \left. \frac{d\ln\epsilon_2}{d \ln r}\right|_{r=R_\star}
\end{equation}
is the logarithmic derivative of the surface harmonic of the distorted star, function of the ellipticity $\epsilon_2$, solution of the Clairaut-Radau differential equation 
\begin{equation}
r\frac{d\eta_2(r)}{dr} + \eta_2^2(r)-\eta_2(r)+6\frac{\rho(r)}{\bar\rho(r)}(\eta_2(r)+1)-6=0
\label{eqn:eta2}
\end{equation}
evaluated at the stellar surface $R_\star$, with the boundary condition $\eta_2(0)=0$ \citep{hejlesen87}. In Eq.\,\ref{eqn:eta2}, $r$ is the radial distance from the core, $\rho(r)$ is the density at the distance $r$, and $\bar\rho(r)$ is the mean density in a sphere of radius $r$. The $k_2$ parameter is thus a measure of the stellar inner density profile, that is to say, the mass distribution between the star’s convective core and radiative envelope. \citet{rosu20a} showed that $d\eta_2/dr$ scales as the density function (see their figure 1)
\begin{equation}
\label{eqn:density_function}
\rho(r) \left(\frac{r}{R_\star}\right)^2 \left(1-\left(\frac{r}{R_\star}\right)^2 \right).
\end{equation}
Measuring the apsidal motion rate hence provides a powerful, independent diagnosis of the density stratification inside stars and offers a unique means to reveal stars’ inner (otherwise hidden) structure. In particular, as pointed out in \citet[their figure 1]{rosu20a} and demonstrated in this paper, $k_2$ is mainly sensitive to the location of the border between the convective core and the radiative envelope, where $d\eta_2/dr$ shows an abrupt change in slope, and to the mixing in this region. This challenging technique has been known for decades but had to wait for long-term spectroscopic monitoring, high-accuracy photometry, and detailed binary stellar evolution codes for its full potential to be exploited. 

In a series of papers, we demonstrated that accurate apsidal motion rates can be derived for double-lined spectroscopic, eclipsing massive binaries, together with accurate fundamental stellar and orbital parameters \citep{rauw16, rosu20b, rosu22a, rosu22b}. Altogether, these stellar parameters enabled the authors to compute tailored stellar evolution models reproducing the properties of the stars such as their mass $M$, radius $R$, effective temperature $T_\text{eff}$, luminosity $L_\text{bol}$, $k_2$, and $\dot\omega$ of the system. \citet{rosu20a, rosu22a} demonstrated that the internal stellar density profiles could be reproduced together with the other parameters provided that enhanced mixing was included in the models, in their case through a large $\alpha_\text{ov}$ and/or high turbulent diffusion. The \texttt{Cl\'es} \citep{scuflaire08} models they computed however suffer from one limitation: They do not include stellar rotation and its ensuing impact on the transport of momentum and chemicals in stars, which makes their models difficult to interpret physically. \citet{claret19} also had to adopt larger $\alpha_\text{ov}$ for the more massive stars of their sample ($4\,M_\odot$) to reconcile the theoretical predictions of $k_2$ with the lower observational $k_2$ determinations. More recently, \citet{fellay24} derived $\alpha_\text{ov}=1.29$ for the massive binary HD\,152248 based on their  non-perturbative approach to derive the apsidal motion rate.

The present work builds upon the works of \citet{rosu20b, rosu20a}. We built bespoke GENeva Evolutionary Code \texttt{GENEC} \citep{eggenberger08} models including both stellar rotation and tides to reproduce the properties of the massive twin binary HD\,152248. The paper is organised as follows. Section\,\ref{sect:observations} summarises the observational properties of HD\,152248. Section \ref{sect:models} introduces the input physics of our \texttt{GENEC} models. Section \ref{sect:single} confronts binary- to single-star \texttt{GENEC} models and investigates the impact of tidal locking on stellar parameter evolution. Section \ref{sect:binary} investigates the influence of input parameters (convective boundary mixing, metallicity, initial helium abundance, initial mass, mass-loss rate, mixing length parameter) on stellar parameter evolution. Section \ref{sect:powr} presents \texttt{PoWR} atmosphere models supplemented to the structures of \texttt{GENEC} models. Section \ref{sect:misalignment} discusses the impact of a possible misalignment of the stellar rotation axes with the orbital plane axis on our interpretation of the apsidal motion in terms of $k_2$. Section \ref{sect:conclusion} provides our conclusions.

\section{Observational properties of HD\,152248\label{sect:observations}}
The very young and rich open cluster NGC\,6231 -- also called the Baby Scorpion Cluster -- to which HD\,152248 belongs is located at the core of the Scorpius OB1 association \citep[][and references therein]{sung98, kuhn17b}. The cluster age ranges between one and seven million years with a small peak around three million years according to the properties of its low-mass stars population \citep{sana07, sung13}, with no obvious global radial age gradient \citep{kuhn17a}. \citet{rosu20a, rosu22b, rosu22a} derived an age estimate of $5.15\pm0.13$, $6.8\pm1.4$, $5.2\pm0.8$, and $9.5\pm0.5$\,Myr for the massive eccentric binaries HD\,152248, CPD-41°7742, HD\,152218, and HD\,152219 in NGC\,6231 based on the study of their apsidal motion. While the first three agree nicely with the estimated cluster age, the last one is more difficult to reconcile with other determinations. 
The metallicity of NGC\,6231 is slightly super-solar: \citet{tadross03, paunzen10} derived [Fe/H]$=0.26\pm0.10$ based on UBVCCD photometric data and \citet{dias21} derived [Fe/H]$=0.11\pm0.07$ based on \textit{Gaia} DR2 data. In this study, we adopt the more recent value of \citet{dias21}. Assuming a proto-solar metallicity $Z_\odot = 0.0142$ \citep{asplund09}, we get $Z_\star = 0.0183^{+0.0032}_{-0.0027}$ for NGC\,6231. The helium content of NGC\,6231 is not well-constrained. \citet{lennon83} derived a slightly super-solar content of $10.98\pm 0.12$\,dex in logarithmic abundance, corresponding to $Y=0.2788$, based on spectra obtained for 24 stars in NGC\,6231 with the 1.9-m telescope and the RGO Cassegrain spectrograph at the South African Astronomical Observatory, while \citet{mathys02} derived a slightly sub-solar value of $10.81\pm0.20$\,dex in logarithmic abundance, corresponding to $Y=0.1885$, using spectra of seven stars located in NGC\,6231 obtained with the ESO Cassegrain Echelle Spectrograph CASPEC mounted on the 3.6-m telescope at La Silla, Chile. Both contents are compatible with the solar value within error bars \citep[$10.93\pm0.01$\,dex in logarithmic abundance, corresponding to $Y=0.2485$,][]{asplund09}. 
The eccentric massive binary HD\,152248 is a twin made of two O7.5 III-II(f) stars. Its fundamental and orbital properties were derived by \citet{rosu20b} based on the combined consistent analysis of spectroscopic and photometric data (Table \ref{table:obs}). 
\setlength{\tabcolsep}{1.5mm}
\begin{table}
\caption{Observational fundamental and orbital parameters of HD\,152248 from \citet{rosu20b}. The system is a twin, hence both stars share the same stellar properties. \label{table:obs}}
\centering
\begin{tabular}{lll}
\hline\hline
Parameter & Symbol (Unit) & Value \\
\hline
\vspace*{-0.3cm}\\
\textit{Stellar}\\
\vspace*{-0.3cm}\\
Mass & $M_\star$ ($M_\odot$) & $29.5^{+0.5}_{-0.4}$ \\
\vspace*{-0.3cm}\\
Radius & $R_\star$ ($R_\odot$) & $15.07^{+0.08}_{-0.12}$ \\
\vspace*{-0.3cm}\\
Effective temperature & $T_{\text{eff},\star}$ (K) & $34\,000 \pm 1000$  \\
\vspace*{-0.3cm}\\
Bolometric luminosity & $L_{\text{bol},\star}$ ($L_\odot$) & $2.73\pm0.32 \times 10^5$ \\
\vspace*{-0.3cm}\\
Apsidal motion constant & $k_{2,\star}$ & $0.0010\pm0.0001$\\
\vspace*{-0.3cm}\\
Surface velocity & $V_{\text{surf},\star}$ (km\,s$^{-1}$) &  $149^{+12}_{-9}$ \\ 
\vspace*{-0.3cm}\\
Mass-loss rate (clumped) & $\dot{M}_\star$ (M$_\odot$\,yr$^{-1}$) & $\le 8.0\times10^{-7}$\\
\vspace*{-0.3cm}\\
Nitrogen abundance & $N/H_\star$ (nb) & $1.24^{+0.98}_{-0.64} \times 10^{-4}$ \\ 
\vspace*{-0.3cm}\\
\textit{Orbital}\\
\vspace*{-0.3cm}\\
Period & $P_\text{orb}$ (d) & $5.816498^{+0.000016}_{-0.000018}$ \\
\vspace*{-0.3cm}\\
Semi-major axis & $a$ ($R_\odot$) & $53.0^{+0.3}_{-0.2}$ \\
\vspace*{-0.3cm}\\
Eccentricity & $e$ & $0.130 \pm 0.002$ \\
\vspace*{-0.3cm}\\
Apsidal motion rate & $\dot\omega$ ($^\circ$\,yr$^{-1}$) & $1.843^{+0.064}_{-0.083}$ \\
\vspace*{-0.3cm}\\
\hspace{0.3cm} Newtonian & $\dot\omega_\text{N}$ ($^\circ$\,yr$^{-1}$) & $1.680^{+0.064}_{-0.083}$ \\
\hspace{0.3cm} General relativistic & $\dot\omega_\text{GR}$ ($^\circ$\,yr$^{-1}$) & $0.163\pm0.001$ \\
\hline
\end{tabular}
\end{table}
\setlength{\tabcolsep}{2.1mm}

\section{\texttt{GENEC} models' input physics\label{sect:models}}
We built single- and binary-star evolution models with \texttt{GENEC} from the onset of hydrogen burning using the stellar physics as in \citet{sciarini26} unless otherwise stated. 

We built two sets of models adopting two different angular momentum transport schemes: purely-hydrodynamic models \citep[calibrated version of][called hydro hereafter]{ekstrom12} and magneto-diffusive models \citep[calibrated Tayler-Spruit dynamo version of][called magnetic hereafter]{eggenberger22}. Magnetic models are calibrated to reproduce surface nitrogen abundances following \citet{sciarini26}. Hydro models account for shear instabilities \citep{zahn92} and meridional circulation \citep{eddington25, sweet50, zahn92, maeder98}, while magnetic models additionally account for the Tayler-Spruit dynamo\footnote{We also computed magnetic models including magneto-rotational instability (MRI)-driven dynamo: It has no impact on the results so we will not consider them further. 
} \citep{spruit02, maeder03, maeder04, maeder05}. The angular momentum transport by meridional circulation is treated as an advective process \citep{ekstrom12} in the hydro models, but is effectively neglected in the magnetic models in which the angular moment transport is treated as a fully diffusive process because magnetic instabilities are strong enough for the star to reach near solid-body rotation during the whole main sequence \citep{maeder05}. In hydro models, the transport of chemicals is treated as a diffusive process where meridional circulation and horizontal turbulence reduce the advection of chemicals. Rotational mixing thus depends on both the angular velocity and the angular velocity gradient of the star \citep{chaboyer92}. In magnetic models, the angular momentum profile is flat inside the star, the shear diffusion coefficient is thus negligible and rotational mixing is dominated by meridional circulation, that is to say, the angular velocity of the star \citep{maeder05, song16, nandal24}. A diffusive approach for chemicals is also applied. 

The binary version of the code currently follows the evolution of the stars until the onset of mass transfer. It accounts for mixing induced by tidal interactions \citep{zahn77,hut81,song13,sciarini24,sciarini26}. For dynamical tides, we used the prescription of \citet[their equation (9)]{sciarini24} that expands the \citet{zahn77} tides formalism for eccentric orbits. The tidal torque is computed as:
\def\restriction#1#2{\mathchoice
              {\setbox1\hbox{${\displaystyle #1}_{\scriptstyle #2}$}
              \restrictionaux{#1}{#2}}
              {\setbox1\hbox{${\textstyle #1}_{\scriptstyle #2}$}
              \restrictionaux{#1}{#2}}
              {\setbox1\hbox{${\scriptstyle #1}_{\scriptscriptstyle #2}$}
              \restrictionaux{#1}{#2}}
              {\setbox1\hbox{${\scriptscriptstyle #1}_{\scriptscriptstyle #2}$}
              \restrictionaux{#1}{#2}}}
\def\restrictionaux#1#2{{#1\,\smash{\vrule height 1.\ht1 depth 1.\dp1}}_{\,#2}} 
\begin{equation}
\begin{aligned}
    &\restriction{\frac{\text{d}}{\text{d}t}\left(I\Omega_{\rm spin}\right)}{\rm Dyn}=\frac{3}{2}\frac{GM^2}{R}E_2q^2\left(\frac{R}{a}\right)^6 \Bigg\{s_{22}^{8/3}\text{sgn}(s_{22})+\\
    &e^2\left(\frac{1}{4}s_{12} ^{8/3}\text{sgn}(s_{12})-5s_{22}^{8/3}\text{sgn}(s_{22})+\frac{49}{4}s_{32}^{8/3}\text{sgn}(s_{32})\right)\Bigg\},
 \end{aligned}
\label{Eq_dyn_tides}
\end{equation}
where $I$ is the moment of inertia of the star, 
\begin{equation}
s_{lm}=(l\Omega_{\rm orb}-m\Omega_{\rm surf})\left(\frac{R^3}{GM}\right)^{1/2}, 
\end{equation}
with $\Omega_{\rm surf}$ and $\Omega_{\rm orb}$ the surface and orbital angular velocities. $E_2$ is a parameter which strongly depends on the ratio of the convective core radius to the stellar radius. We computed it using the prescription of \citet{qin18} for main-sequence stars: 
\begin{equation}
E_2 = 10^{-0.42}\left(\frac{R_{\rm conv}}{R}\right)^{7/5}.
\end{equation}
For equilibrium tides of sub-surface convective zones, we used the prescription of \citet[their equations (8) and (9)]{sciarini26}, adapted from the formalism of \citet{hut81} for the case of small sub-surface convective zones, inspired by the treatment by \citet{fragos23}. In case of eccentric orbits, the angular velocity equation generalises to:
\begin{equation}
\begin{aligned}
    \restriction{\frac{\text{d}}{\text{d}t}\left(I\Omega_{\rm surf}\right)}{\rm Eq}=~ 
    &3\frac{k_2}{\tau_{\rm conv}}f_{\rm conv}\frac{I_{\rm conv.reg.}}{I_*}MR^2q^2\left(\frac{R}{a}\right)^6 \frac{\Omega_{\rm orb}}{(1-e^2)^{6}} \\
    &\left(f_2-(1-e^2)^{3/2}f_5\frac{\Omega_{\rm surf}}{\Omega_{\rm orb}}\right),
\end{aligned}
\label{Eq_eq_tides}
\end{equation}
where $\tau_{\rm conv}$ is the convective turnover timescale, $f_{\rm conv}$ is a factor which reduces the strength of the tides when the tidal pumping timescale gets smaller than $\tau_{\rm conv}$, $I_{\rm conv.reg.}$ is the moment of inertia of the convective region, and $f_2$ and $f_5$ are functions of the eccentricity, defined in \citet{hut81}.

For single-star models, we tested different initial velocities of the star, from 0.1 to 0.3, set through the parameter $v_\text{ini}/v_\text{cr}$, where $v_\text{cr}$ is the critical velocity of the star (Sect.\,\ref{sect:single}). For binary models, we initialised the system synchronised with the orbital motion and assumed the orbital period fixed to the observational value throughout the evolution. We defer to Sects \ref{subsect:not_synchronised} and \ref{subsect:orbital_period_not_fixed} for a discussion of the impact of not initialising the system synchronised as well as leaving the orbital period vary during the evolution, respectively.  The companion is assumed to have the same mass as the primary star throughout the evolution (i.e. mass-loss rate is also applied to the companion star through a reduction of its mass at each time step).  

Overshooting is introduced in the models using the step-overshoot parameter $\alpha_\text{ov}$ that we varied from 0.5 \citep[suggested by][]{higgins19} to 1.4 (Sect.\,\ref{subsect:alpha_ov}). We adopted a metallicity $Z=0.0183$ and tested $Z=0.0156$ and 0.0215 (see Sect.\,\ref{sect:introduction}, Sect.\,\ref{subsect:Z}). Given that the helium content of the cluster is unconstrained, we adopted the solar value corrected by the super-solar metallicity: $Y_\odot=0.2696^{+0.0037}_{-0.0031}$ at $Z=0.0183^{+0.0032}_{-0.0027}$, corresponding to $X=0.7121^{-0.0069}_{+0.0058}$. We explored the impact of an enhanced initial helium content by increasing the values of $Y_\odot$ by 0.05 and 0.10 (Sect.\,\ref{subsect:Z}). Outer convective zones are treated following the mixing length theory with a mixing length parameter $\alpha_\text{MLT} = l/H_\text{P} = 1.6$ for stars of masses 1.25 to 40\,$M_\odot$, where $l$ is the mixing length \citep{ekstrom12}. Following the study of \citet{joyce23}, we varied $\alpha_\text{MLT}$ from 0.1 to 3.0 (Sect.\,\ref{subsect:alpha_MLT}). We adopted the Schwarzschild convection criterion\footnote{We also computed models with the Ledoux criterion and observed no difference compared to models with the Schwarzschild criterion, which is expected during the main sequence as models have a large overshooting that prevents semiconvection regions to develop above the core \citep{kaiser20}.}.

Stellar wind mass-loss is included through the prescription of \citet{krticka24}. This choice is motivated by the facts that, on the one hand, the \citet{vink01} prescription has been shown to overpredict the mass-loss rate of massive stars by a factor $\sim$3 \citep{brands22} owing to the fact that massive star winds are clumped \citep{surlan12}, and, on the other hand, the \citet{krticka18} prescription \citep[which][is an extension of for B supergiants]{krticka24} is found to better reproduce observations of massive stars than the \citet{bjorklund21}  prescription in their common validity range \citep[see figures 2 and 12 of][respectively]{hawcroft24,brands22}. \citet{krticka18, krticka24} point out that if micro-clumping (commonly simply called clumping) happens to start close to the sonic point, that is to say, the stellar photosphere, the theoretical mass-loss predictions would be underestimated by a factor $C_\text{c}^{0.25}$, where $C_\text{c} = 1/f_\infty$ is the clumping factor, $f_\infty$ being the volume filling factor. \citet{krticka17} derived $C_\text{c} = 8$ for their sample of O stars in the Milky Way and \citet{krticka18} derived $C_\text{c} = 9$ for their sample of O stars in the Large and Small Magellanic Clouds. In both cases, $C_\text{c}^{0.25} = 1.7$. \citet{hawcroft24} showed that late O supergiants have $C_\text{c}$ systematically lower than 10, while O giants have an intermediate $C_\text{c}$ of $10-25$, earlier type-stars having the largest values. Given that HD\,152248 is an intermediate-late O star between luminosity classes III and II (Sect.\,\ref{sect:observations}), if we adopt $C_\text{c} = 10$ -- in agreement with the \texttt{CMFGEN} spectroscopic analysis performed by \citet{rosu20b} to derive the clumped mass-loss rate --, the scaling factor amounts to $10^{0.25} = 1.78$, larger than the 1.7 upper limit expected for most massive stars but still reasonable (J. Krti\v{c}ka, priv. com.). Therefore, we also computed models with the mass-loss rate prescription of \citet{krticka24} to which a scaling factor $\xi$ of 1.78 was applied.

\section{Single- versus binary-star models: the impact of tidal locking\label{sect:single}}

\subsection{Single-star models: velocity evolution from stellar evolution}
We investigated the impact of the initial velocity $v_\text{ini}$, all other parameters fixed, on the evolution of the star. In particular, we focused on the stellar parameters of our models having a radius equal to the observed one $R_\star$ within the error bars. Indeed, if we want to use Eq.\,\eqref{eqn:omegadot} as an additional constraint for the stars through $k_{2,\star}$, we see that a model that does not accurately reproduces $R_\star$ will lead to spurious results, given the dependence of $\dot\omega_\text{N}$ in $R_\star^5$. It also justifies our choice to present the evolution of models properties as a function of radius and not of stellar age. The models have $\alpha_\text{ov}=0.5$, $Z=0.0183$, and $Y_\odot$. We set the initial mass of the models to $30.2\,M_\odot$, such that most of our models have $M(R_\star) = M_\star$.

\begin{table*}
\caption{Influence of an increase in the value of inputs parameters in \texttt{GENEC} ($v_\text{ini}$, $\alpha_\text{ov}$, $Z$, $Y$, $M_\text{ini}$, and clumping $\xi$) on stellar parameters given in Col.\,1 for hydro (Cols 2-7) and magnetic (Cols 8-13) models. Parameters are compared at fixed radius equal to $R_\star$. See main text in Sects \ref{sect:single} and \ref{sect:binary}. \label{table:influence}}
\centering
\begin{tabular}{lccccccccccccc}
\hline\hline
& \multicolumn{6}{c}{Hydro models} && \multicolumn{6}{c}{Magnetic models}   \\
Impact on & \multicolumn{6}{c}{Increase in} && \multicolumn{6}{c}{Increase in}  \\
Parameter & $v_\text{ini}^a$ & $\alpha_\text{ov}$ & $Z$ & $Y$ & $M_\text{ini}$ &$\xi$ & & $v_\text{ini}^a$ & $\alpha_\text{ov}$  & $Z$ & $Y$ & $M_\text{ini}$ & $\xi$   \\
\vspace*{-0.3cm}\\
\hline
Age & $\rightarrow^*$  & $\uparrow$ & $\downarrow$ & $\downarrow$ & $\downarrow$ & $\downarrow$ & & $\uparrow$ & $\uparrow$ & $\downarrow$ & $\downarrow$  & $\downarrow$ & $\downarrow$ \\
$M$ & $\rightarrow^*$ & $\downarrow$  & $\downarrow$ & $\rightarrow^*$ &$\uparrow$ & $\uparrow$ &   & $\downarrow$ & $\downarrow$ & $\downarrow$ & $\rightarrow^*$ &  $\uparrow$ & $\uparrow$ \\
$T_\text{eff}$ & $\rightarrow$ &$\uparrow$  & $\downarrow$ & $\uparrow$ & $\uparrow$ & $\uparrow$  &&  $\uparrow$ & $\uparrow$ & $\downarrow$ & $\uparrow$ & $\uparrow$ & $\uparrow$   \\
$L_\text{bol}$ & $\rightarrow$ & $\uparrow$  & $\downarrow$ & $\uparrow$ & $\uparrow$ & $\uparrow$  && $\uparrow$ & $\uparrow$ & $\downarrow$ & $\uparrow$ & $\uparrow$ & $\uparrow$ \\
$k_2$ & $\downarrow$ & $\downarrow$ & $\downarrow$ & $\downarrow$ & $\uparrow^*$ & $\rightarrow^*$ && $\downarrow$ & $\downarrow$ & $\downarrow$ & $\downarrow$ &$\uparrow^*$ & $\rightarrow^*$  \\
$V_\text{surf}$ & $\uparrow$ & $\uparrow$ & $\downarrow$/$\uparrow^*$ & $\rightarrow^*$ & $\rightarrow^*$ & $\downarrow^*$ &&  $\uparrow$ & $\uparrow$ & $\downarrow$/$\uparrow^*$ & $\rightarrow^*$ & $\rightarrow^*$ & $\downarrow^*$ \\
$N/H$ & $\uparrow$ & $\rightarrow^*$/$\uparrow$  & $\uparrow$ & $\uparrow$ & $\rightarrow^*$ & $\uparrow$ && $\uparrow$ & $\uparrow$ & $\uparrow$ & $\uparrow$ & $\rightarrow^*$ & $\uparrow$ \\
$\dot{M}$ & $\uparrow$ &$\uparrow$ &$\uparrow$ & $\uparrow$ & $\uparrow$ & $\uparrow$ && $\uparrow$ &$\uparrow$ & $\uparrow$ & $\uparrow$ & $\uparrow$ & $\uparrow$ \\
\hline
\end{tabular}
\begin{tablenotes}
\item \textbf{Notes.} $^a$Only valid for single-star models as binary-star models are initialised synchronised. $\uparrow$ and $\downarrow$ mean respectively an increase and decrease in the parameter at $R=R_\star$. $\rightarrow$ means the trend is constant or not clear and depends on other parameters. $^*$means the change is not significant. If only one arrow is given, it applies to both single- and binary-star models; If two arrows are given the first (resp. second) one refers to single- (resp. binary-) star models.
\end{tablenotes}
\end{table*}

 \begin{figure}[t]
 \centering
 \includegraphics[width=\linewidth]{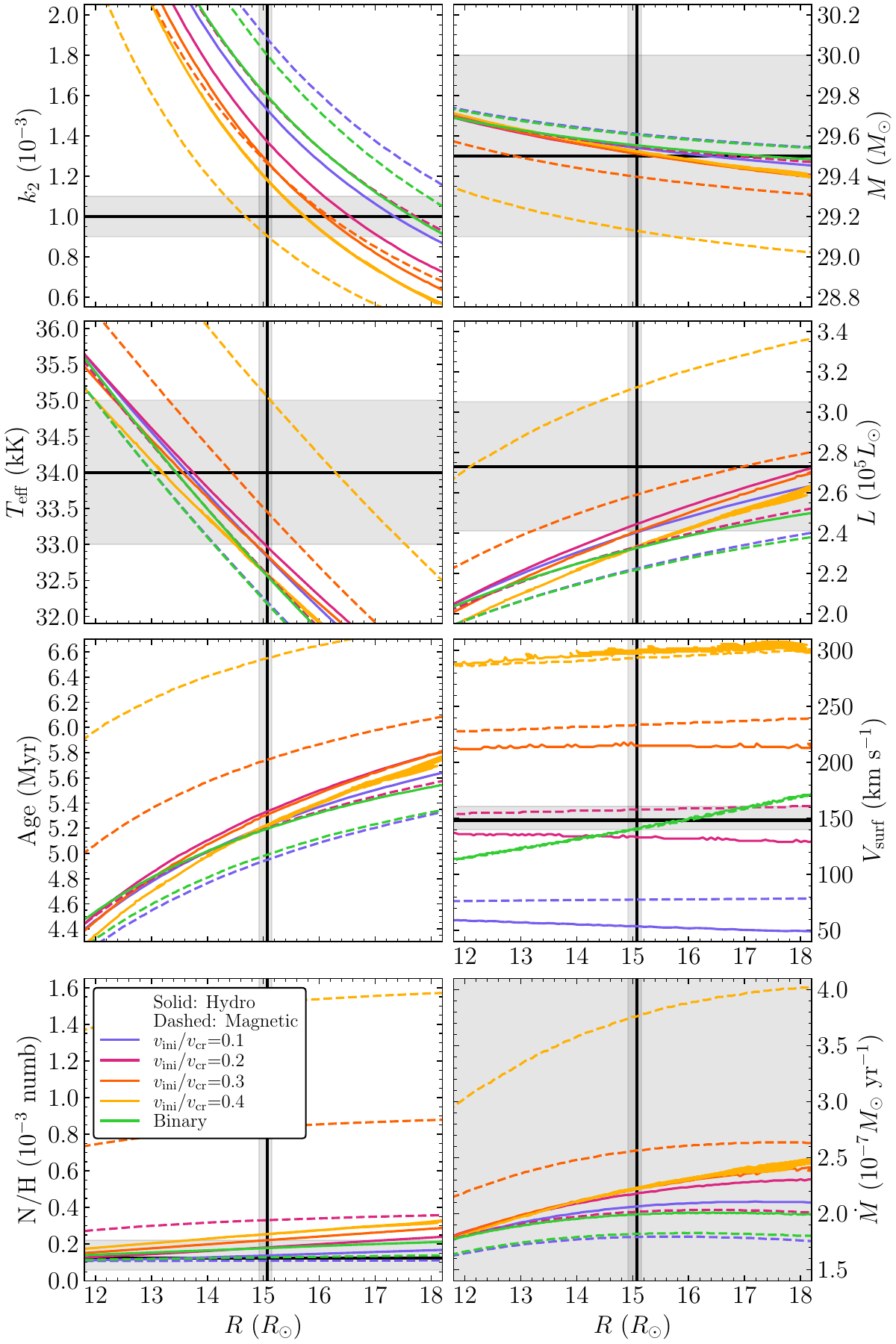}
 \caption{Evolution of stellar parameters with stellar radius for hydro (plain lines) and magnetic (dashed lines) single-star models for different values of $v_\text{ini}$ (colours). The models have $M_\text{ini} = 30.2M_\odot$, $\alpha_\text{ov} = 0.5$, $Z=0.0183$, $Y_\odot$, and \citet{krticka24} mass-loss rate. Binary-star models initialised synchronised, all other parameters identical, are shown for comparison. Observational values and their error bars are shown (plain black line and grey shaded area). \label{fig:vinit}}
 \end{figure}
 
\begin{figure}[h!]
\includegraphics[width=\linewidth]{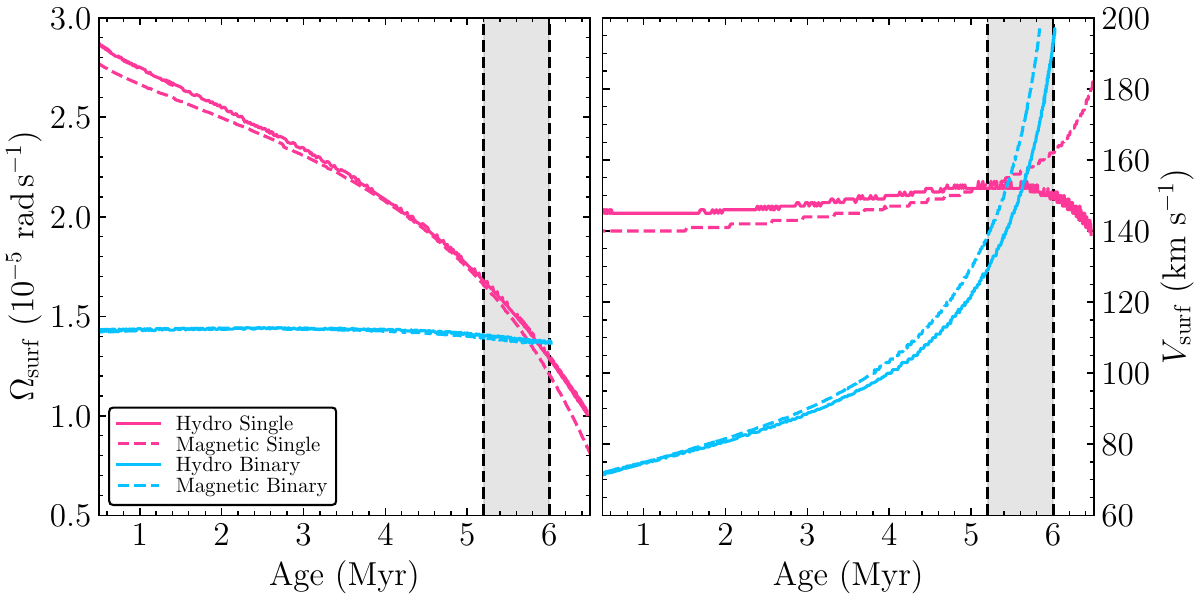}
\caption{Evolution of the surface angular velocity $\Omega_\text{surf}$ (left panel) and surface velocity $V_\text{surf}$ (right panel) with stellar age for single and binary-star models. The models have $M_\text{ini} = 30.2M_\odot$, $\alpha_\text{ov} = 0.8$, $Z=0.0183$, $Y_\odot$, and \citet{krticka24} mass-loss rate. Models that reproduce $R_\star$ are in the gray boxes. \label{fig:angularmomentum}}
\end{figure}

The impact of $v_\text{ini}$ on stellar evolution is shown in Fig.\,\ref{fig:vinit}. The main dependencies are summarised in Table\,\ref{table:influence}.  A larger $v_\text{ini}$ triggers more efficient internal mixing responsible for the transport of CNO-processed material to the surface, which leads to a larger surface nitrogen abundance $N/H$. Models with larger $v_\text{ini}$ are older when they reach $R_\star$ thus have lost more mass than their lower-$v_\text{ini}$ counterpart. 
The largest the $v_\text{ini}$, the smallest the mass of the star at $R_\star$ due to the larger luminosity that induces a larger mass-loss rate. The effect is more pronounced in magnetic models than in hydro models as the mixing induced by rotation is more efficient in magnetic models as it only depends on the angular velocity content of the star.
Magnetic models with larger $v_\text{ini}$ evolve towards bluer parts of the Hertzsprung-Russell diagram, are older when they reach $R_\star$, and consequently have smaller $k_2$ values. On the opposite, there is no clear trend between $v_\text{ini}$ and age for hydro models: They all reach $R_\star$ at the same age $\pm$\,$0.1$\,Myr. It is the consequence of two counteracting effects. On the one hand, models with larger $v_\text{ini}$ are more mixed so their radius increases less fast. On the other hand, models with larger $v_\text{ini}$ get closer to the critical velocity; hydrostatic effects start to play an important role. The observed decrease in $k_2$ with increasing $v_\text{ini}$ is thus the direct consequence of larger mixing in the models. 

Only the hydro model with $v_\text{ini}/v_\text{cr} = 0.2$ reproduces $M_\star, R_\star, T_{\text{eff},\star}, L_{\text{bol},\star}, V_{\text{surf},\star}, N/H_\star$, and $\dot{M}_\star$. The magnetic model with $v_\text{ini}/v_\text{cr} = 0.2$ that reproduces $M_\star, R_\star$, and $V_{\text{surf},\star}$ predicts too low values for $T_\text{eff}$ and $L_\text{bol}$, and a too large value for $N/H$. Yet, all models predict a too large $k_2$-value. 

\subsection{Binary-star models: velocity evolution fixed by tides}
Two binary-star models computed with the same parameters as single-star models are shown in Fig.\,\ref{fig:vinit}. The models are initialised synchronised with the orbital motion, meaning that the star's angular velocity is fixed by its orbital angular velocity. The binary-star models do not reproduce the stellar properties as well as their single-star counterparts. Indeed, the $k_2$ of binary models are even larger, while both $T_\text{eff}$ and $L$ are significantly underestimated. 

If the single-star models are apparently better at reproducing the current observed properties of the stars, it is because we chose $v_\text{ini}$ so that it is the case, that is to say, so that $V_\text{surf}(R_\star) = V_{\text{surf},\star}$. A single star sees its $\Omega_\text{surf}$ decrease with time as a consequence of its surface breaking due to its radius increasing and of the mass lost due to winds carrying out angular momentum. Consequently, its $V_\text{surf}$ is roughly constant over time (Fig.\,\ref{fig:angularmomentum}). In reality, the star in a binary system has a companion; Its $\Omega_\text{surf}$ is thus fixed by the orbital motion and the binary is pseudo-synchronised. The star's radius increases with time, so does its $V_\text{surf}$ (Fig.\,\ref{fig:angularmomentum}). Even though both single- and binary-star models predict the same $\Omega_\text{surf}$ and $V_\text{surf}$ at $R_\star$, both quantities were lower during the whole evolution of the star in a binary. Therefore, the star in a binary has had less mixing throughout its evolution, and has a larger $k_2$ at $R= R_\star$. $N/H$ in binary-star models is in between $N/H$ of single-star models with $v_\text{ini}/v_\text{cr}=0.1$ and 0.2, consistent with the evolution of velocities. 

It illustrates the importance of accounting for binary interactions in stellar models in short-period systems, as tides can suppress rotational mixing and drastically change the evolution of a star compared to the single-star case \citep[see][for a thorough discussion of the impact of tides in short-period systems]{sciarini26}. As of now, we will consider binary-star models only. 

\subsubsection{Binary-star models not initialised synchronised\label{subsect:not_synchronised}}
We investigated the impact of not initialising the system synchronised. We computed models adopting $v_\text{ini}/v_\text{cr} =0.1$ and $0.2$ and compared their evolution to synchronised models (see Fig.\,\ref{fig:notsynchro}). Past the synchronisation of the initially not synchronised models that takes $\sim$0.5\,Myr, the evolution in terms of $M$, $T_\text{eff}$, $L$, and $\dot{M}$ is identical in models with $v_\text{ini}/v_\text{cr}=0.1$ and 0.2 and synchronised. At $R=R_\star$, all models have the same age. However, hydro models with $v_\text{ini}/v_\text{cr}=0.2$ have a surface $N/H$ twice that of the synchronised models. It comes from the short initial phase of synchronisation during which stars with $v_\text{ini}/v_\text{cr}=0.2$ significantly enrich their surface due to rotational mixing proportional to both the angular velocity and its gradient. 
The evolution of $k_2$ is much less impacted by the choice of initial conditions for $v_\text{ini}$ than other model parameters (see Sect.\,\ref{sect:binary}). We conclude that initialising the system synchronised is a valid assumption.

\subsubsection{Binary-star models without fixing the orbital period \label{subsect:orbital_period_not_fixed}}
We tested the impact of leaving the orbital period of the system vary during the evolution under the influence of tides, setting the initial period to the observed value. The fundamental parameters of the star ($M, T_\text{eff}, L$) are not affected, but $\Omega_\text{surf}$ and hence $V_\text{surf}$ are 6\% lower when $R = R_\star$ due to the larger orbital period (see Fig.\,\ref{fig:notsynchro}). Hence, $k_2$ is slightly but not significantly larger. We conclude that fixing the orbital period to the observed value throughout the evolution of the star is a valid assumption.


\begin{figure}[t]
\centering
\includegraphics[width=\linewidth]{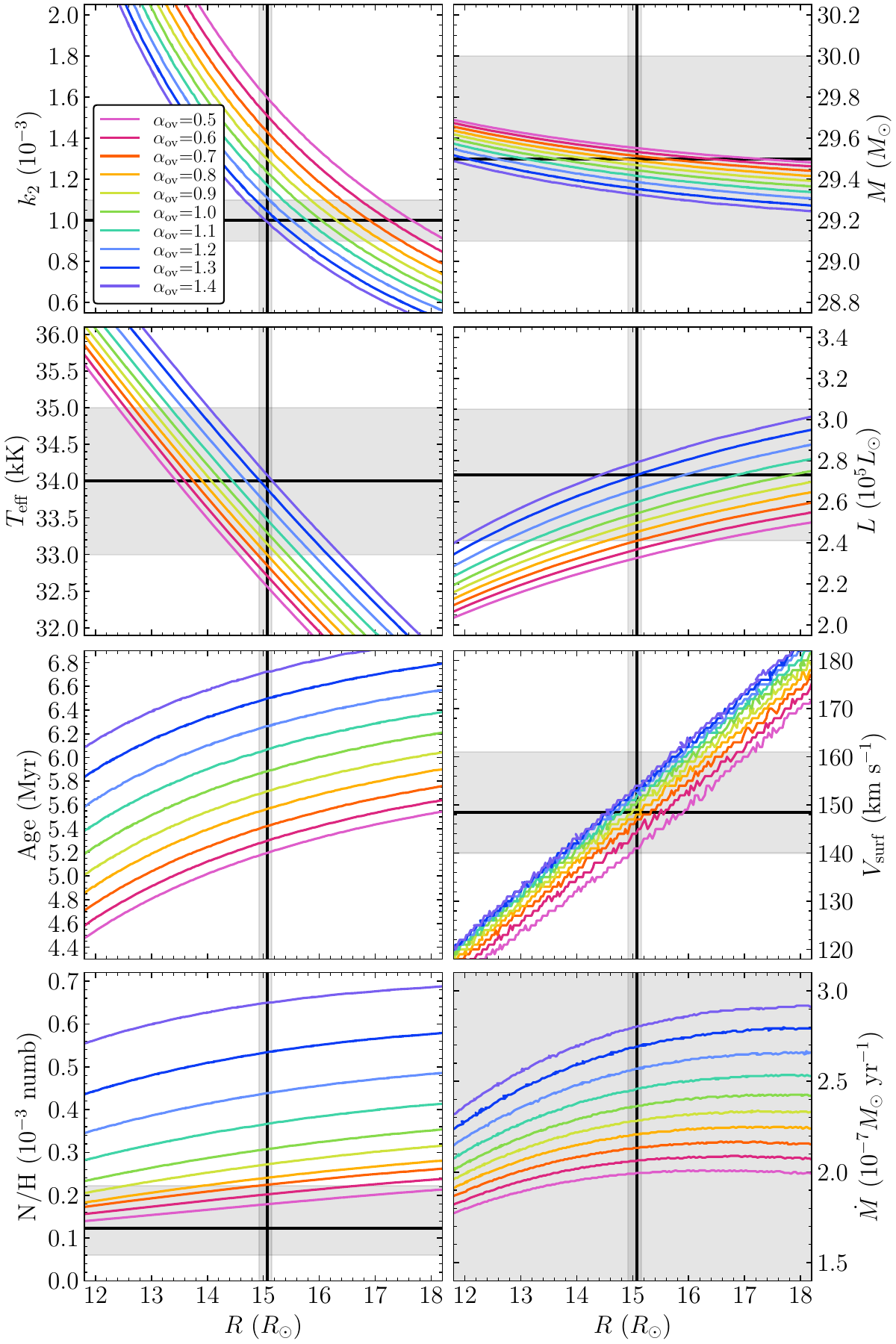}
\caption{Evolution of stellar parameters with stellar radius for hydro binary-star models for different values of $\alpha_\text{ov}$. Models have $M_\text{ini} = 30.2M_\odot$, $Z=0.0183$, $Y_\odot$, and \citet{krticka24} mass-loss rate. Observational values and their error bars are shown (plain black line and grey shaded area). \label{fig:alpha_ov}}
\end{figure}
\section{Binary-star models: parameter space exploration\label{sect:binary}}
We investigated the impact of modifying one input parameter at a time in binary-star models, all other parameters fixed, on the evolution of the star. The main dependencies are explained in the following sections and summarised in Table\,\ref{table:influence}. Our reference model has $\alpha_\text{ov}=0.5$, $Z=0.0183$, $Y_\odot$, \citet{krticka24} mass-loss rate, and $\alpha_\text{MLT} = 1.6$. We set $M_\text{ini}$ of the models to $30.2\,M_\odot$, so that most of our models have $M(R_\star) = M_\star$. 

\subsection{Overshooting parameter $\alpha_\text{ov}$ \label{subsect:alpha_ov}}
The impact of $\alpha_\text{ov}$ on stellar evolution is shown in Fig.\,\ref{fig:alpha_ov} for hydro models and in Fig.\,\ref{fig:alpha_ov_mag} for magnetic models. Increasing $\alpha_\text{ov}$ of a model gives rise to a blueward evolution in the Hertzsprung-Russell diagram and an extended main-sequence lifetime. Consequently, when the star reaches $R_\star$, it has larger effective temperature and luminosity and has evolved for a longer time, thus shows a higher density contrast between its dense core and extended envelope, that is to say, it has a lower $k_2$.

\begin{figure}
\includegraphics[width=\linewidth]{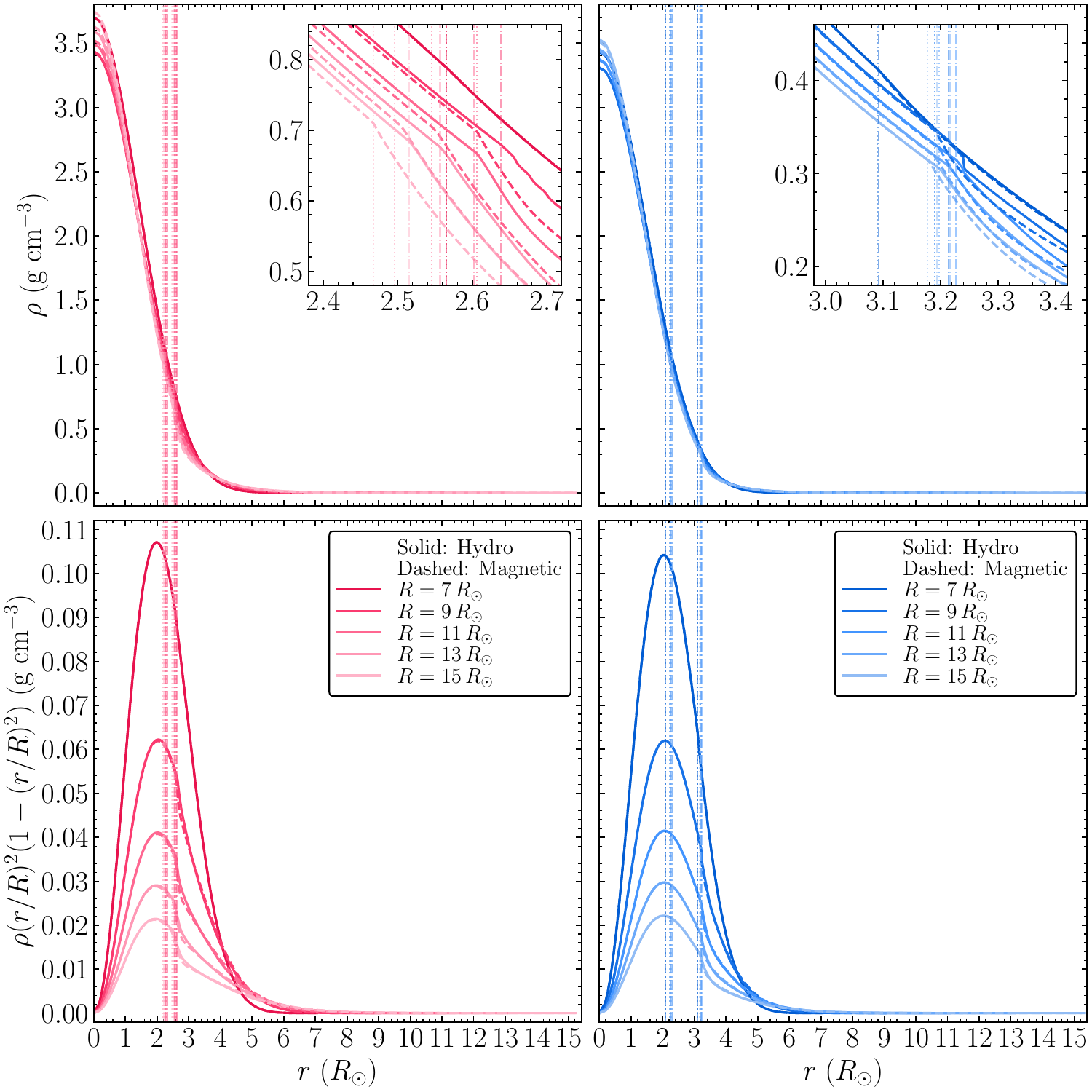}
\caption{Density stratification (\textit{top panels}) and density function (Eq.\,\eqref{eqn:density_function}, \textit{bottom panels}) in binary-star models as a function of the radial distance $r$ in the star. Models have $\alpha_\text{ov}=0.5$ (\textit{left panels}) or $\alpha_\text{ov}=1.4$ (\textit{right panels}), $Z=0.0183$, $Y_\odot$, and \citet{krticka24} mass-loss rate. Models have increasing age from dark to light shades (the lightest shade models have $R_\star$). The location of the border of the convective core and the overshooting region are indicated by vertical lines (dotted-dashed and dotted for hydro and magnetic models, respectively). \label{fig:density_radius_time}}
\end{figure}
\begin{figure*}
\includegraphics[width=0.49\linewidth]{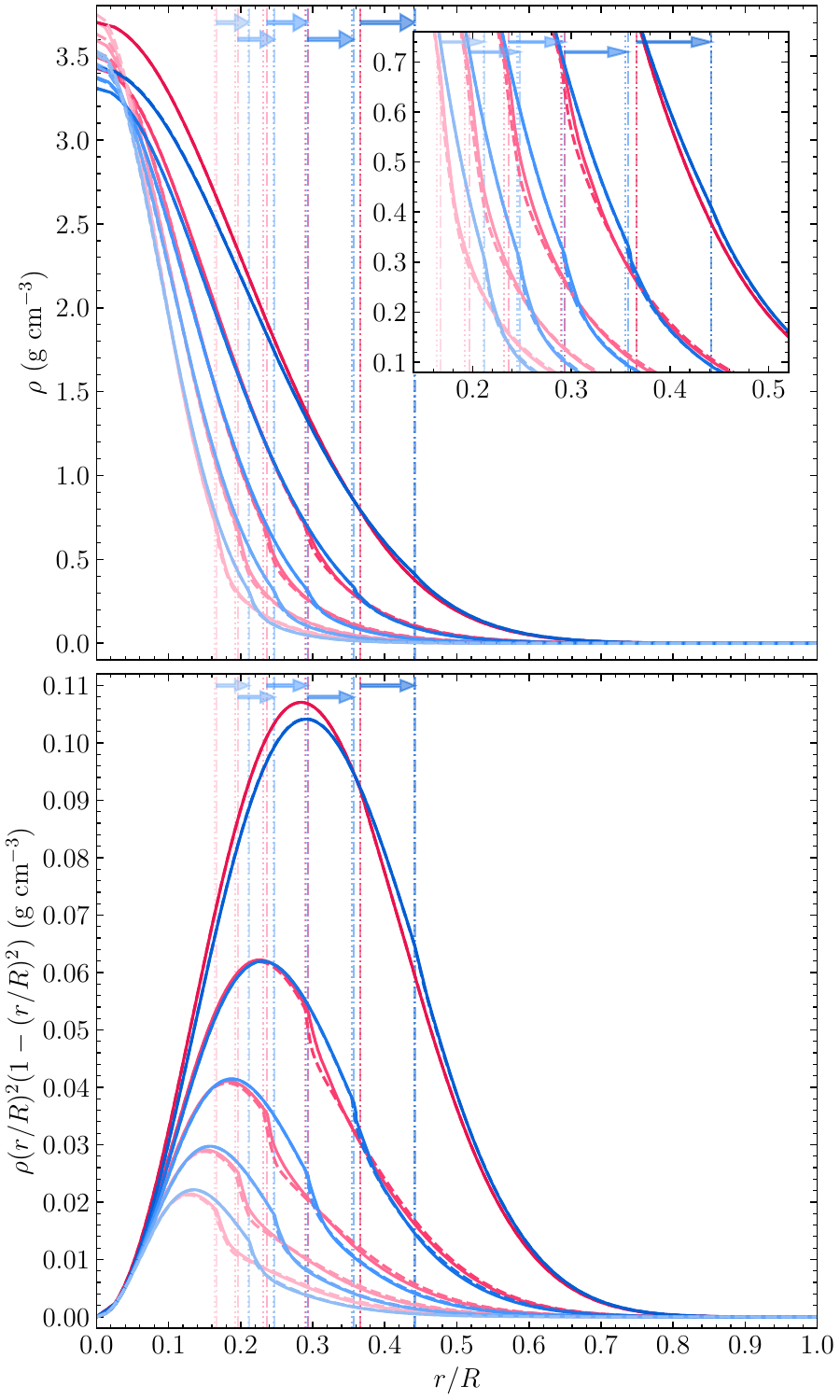}
\includegraphics[width=0.49\linewidth]{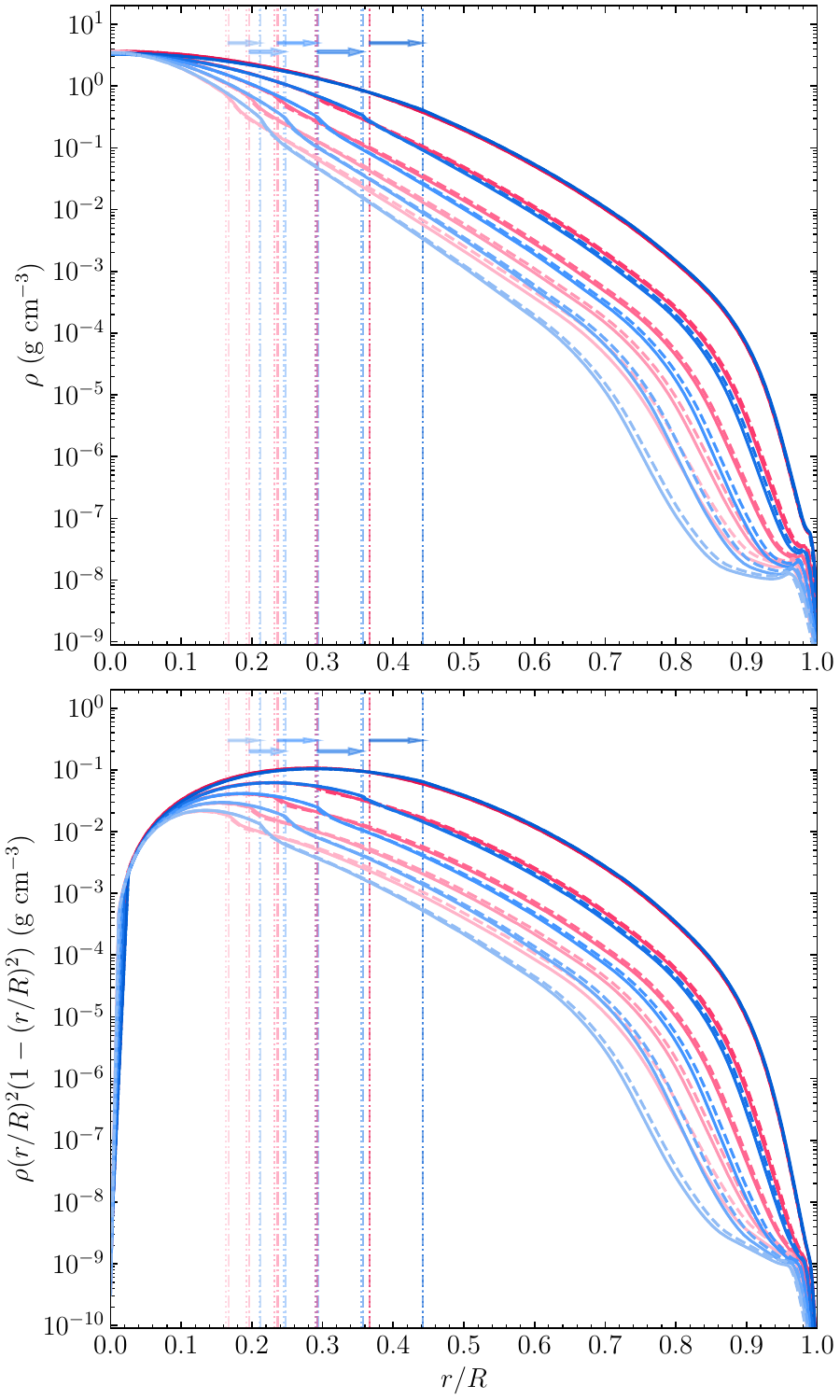}
\caption{Density stratification (\textit{top panels}) and density function (Eq.\,\eqref{eqn:density_function}, \textit{bottom panels}) in binary-star models as a function of the normalised radial distance $r/R$ in the star for the same models as in Fig.\,\ref{fig:density_radius_time}. \textit{Left/Right panels:} Linear or log scale are adopted to better highlight the central and envelope regions, respectively. The location of the border of the overshooting region is indicated by vertical lines (dotted-dashed and dotted for hydro and magnetic models, respectively). Horizontal arrows symbolise the extension of the border of the overshooting region from models with $\alpha_\text{ov}=0.5$ to those with $\alpha_\text{ov}=1.4$.  \label{fig:density_radius_time_stacked}}
 \end{figure*}

Overshooting acts mainly at the junction between the convective core and the radiative envelope of the star, and has no direct effect on the external layers of the star. The mass-loss rate is therefore unchanged at a given time and the mass is not directly affected by a change in overshooting. The slight decrease in mass at $R_\star$ with overshooting only results from the older age of the star when it reaches $R_\star$. When $R=R_\star$, all models are slightly de-synchronised because dynamical tides decreased in intensity while equilibrium tides are not yet strong enough to re-synchronise the system \citep[see Appendix B in][]{sciarini26}. Models with larger $\alpha_\text{ov}$ are better synchronised because their $E_2$-coefficient appearing in the synchronisation timescale is larger \citep[see equation 9 in][]{sciarini24} due to their larger core. Therefore, models with larger $\alpha_\text{ov}$ have slightly larger $V_\text{surf}$. 
$N/H$ is not significantly nor directly affected by a change in overshooting: It slightly increases with overshooting as a consequence of the smaller distance between the border of the extended convective core and the stellar surface, facilitating the transport of nitrogen to the surface. 

The difference in $k_2$ between models directly follows from the different core-envelope density constrats. It is illustrated by the density profiles inside stars as well as by the density function (Eq.\,\eqref{eqn:density_function}) for models with $\alpha_\text{ov} = 0.5$ and 1.4 as a function of $r$ in Fig.\,\ref{fig:density_radius_time} and of $r/R$, where $R$ is the radius of the model, in Fig.\,\ref{fig:density_radius_time_stacked}. As a star evolves, its core contracts and envelope expands, thus the core density $\rho_\text{c}$ increases while the envelope density decreases. At $\alpha_\text{ov} = 0.5$, $\rho_\text{c}$ is systematically larger in magnetic models than in hydro models, while at $\alpha_\text{ov}=1.4$ the difference is not significant. 
The density profile and density function show a small hook at the junction between the overshooting region and the radiative envelope. This hook is located at a distance from the core of the star of $r$\,$\sim$\,$2.6\,R_\odot$ and $\sim$\,$3.2\,R_\odot$ when $R=R_\star$ in models with $\alpha_\text{ov} = 0.5$ and 1.4, respectively, encompassing masses $M_r$ of $\sim$\,$17.5\,M_\odot$ and $\sim$\,$23.5\,M_\odot$, respectively. This junction reaches further down into the core of the star as the star evolves. The hook in hydro models is systematically located at a slightly larger $r$ than in magnetic models (the difference is more pronounced at lower $\alpha_\text{ov}$) while the density in the envelope is lower. It explains why hydro models slightly better reproduce the observed $k_2$ as the density contrast is more pronounced. The density function shows a lower peak value at larger $R$, hence the lower $k_2$. After the hook, models with $\alpha_\text{ov} = 1.4$ show a sharper decrease in the density function than models with $\alpha_\text{ov} = 0.5$: The contrast in the density function between the core and the envelope is larger at larger $\alpha_\text{ov}$, hence the lower $k_2$. 
Only hydro models with $\alpha_\text{ov} \ge 1.2$ and magnetic models with $\alpha_\text{ov} \ge 1.3$ reproduce all stellar parameters simultaneously, including $k_2$. Yet, these models overpredict $N/H$.

\subsection{Metallicity $Z$ and initial helium content $Y$ \label{subsect:Z}}
\begin{figure}
\centering
\includegraphics[width=\linewidth]{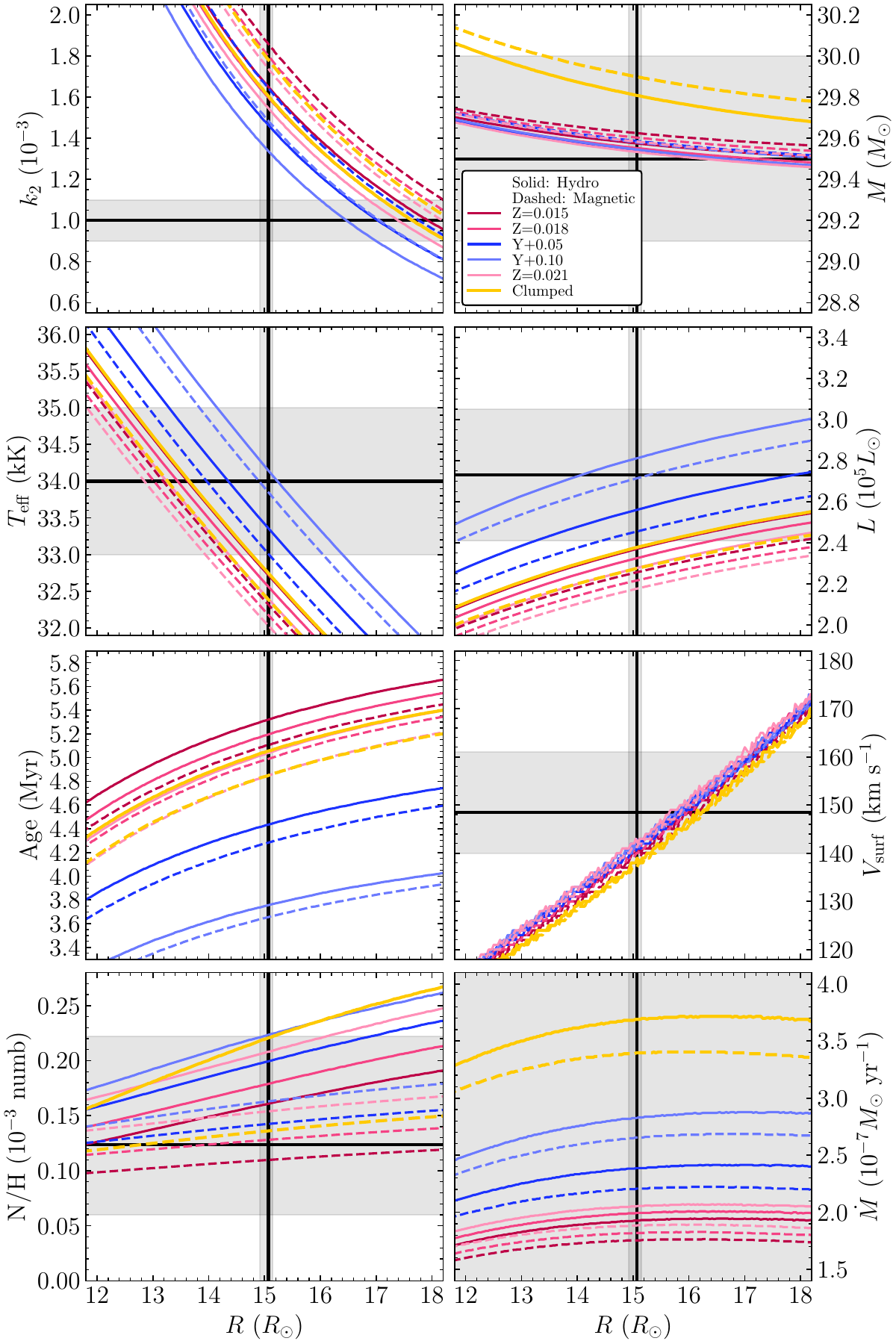}
\caption{Evolution of stellar parameters with stellar radius for hydro (plain lines) and magnetic (dashed lines) binary-star models of $M_\text{ini} = 30.2M_\odot$, $\alpha_\text{ov}=0.5$, and \citet{krticka24} mass-loss rate for different values of $Z$ and $Y$. A clumped model with $M_\text{ini} = 31.0M_\odot$, $\alpha_\text{ov}=0.5$, $Z=0.0183$, $Y_\odot$, and \citet{krticka24} mass-loss rate multiplied by $\xi=1.78$ is also shown. Observational values and their error bars are shown (plain black line and grey shaded area). \label{fig:zhe}}
 \end{figure}

\begin{figure*}
\includegraphics[width=0.49\linewidth]{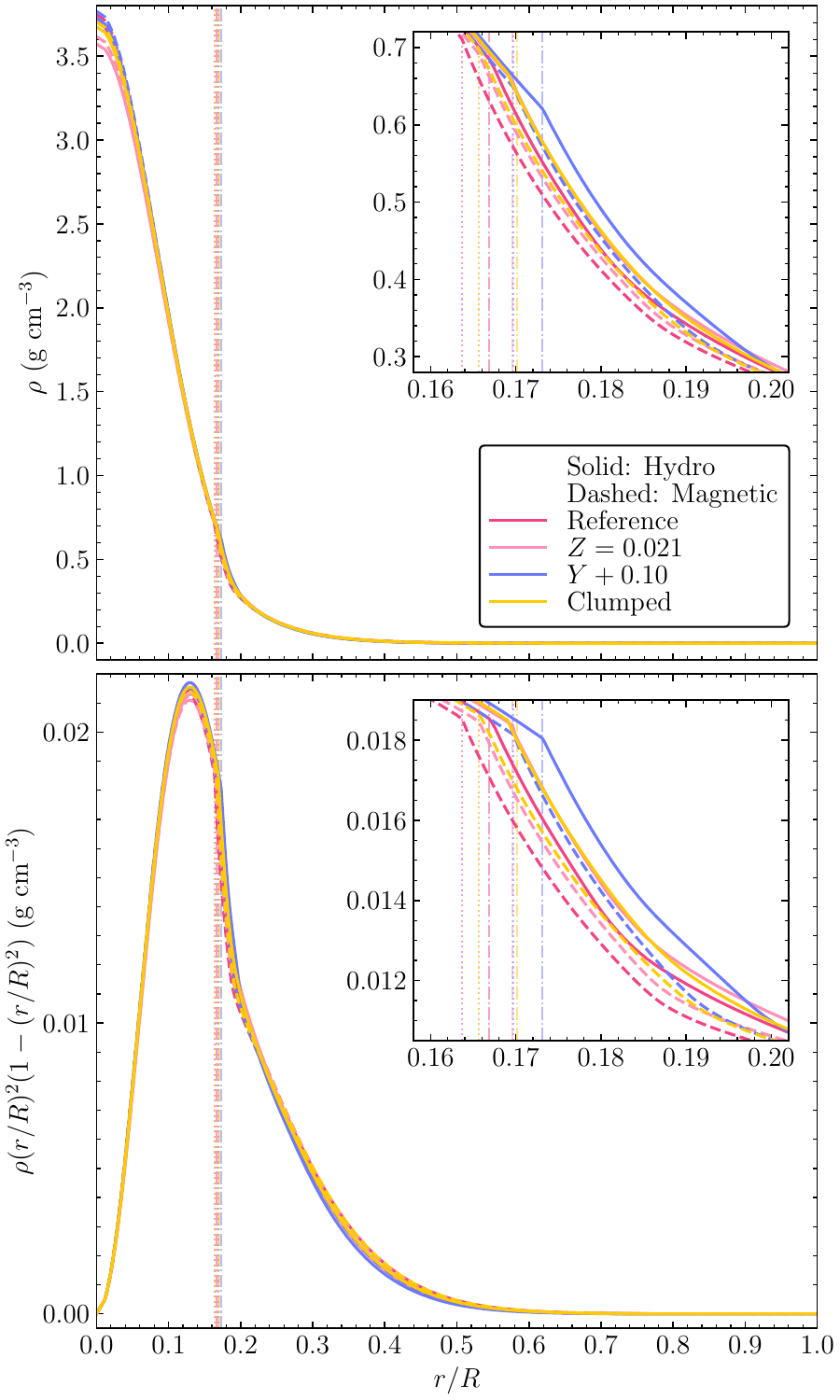}
\includegraphics[width=0.49\linewidth]{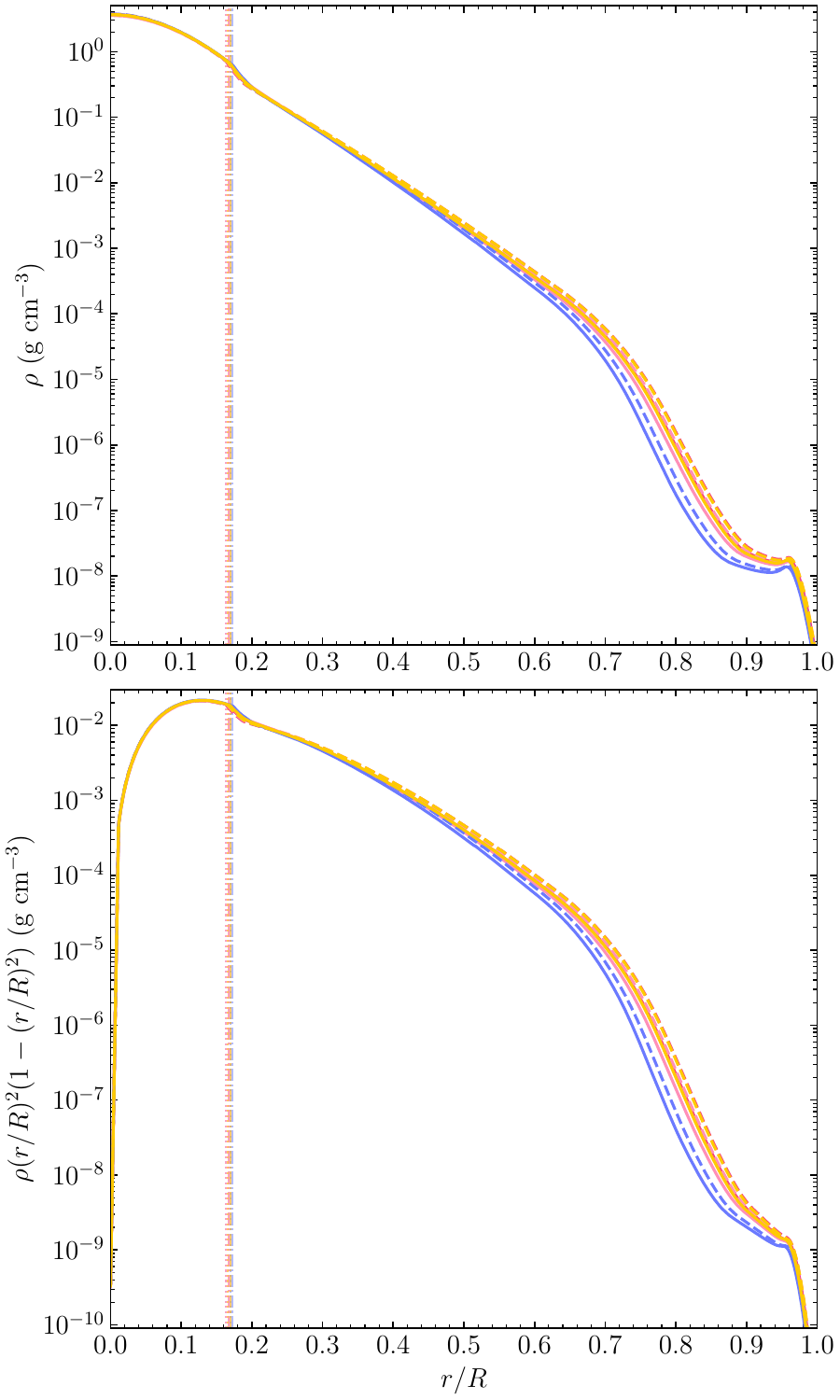}
\caption{Density stratification (\textit{top panels}) and density function (Eq.\eqref{eqn:density_function}, \textit{bottom panels}) in binary-star models with $R=R_\star$  as a function of the normalised radial distance $r/R$ in the star. The reference model (fuchsia) has $\alpha_\text{ov}=0.5$, $Z=0.0183$, $Y_\odot$, and \citet{krticka24} mass-loss rate. Models with $Z=0.0215$, $Y_\odot+0.10$, and \citet{krticka24} mass-loss rate multiplied by 1.78, all other parameters identical, are shown (pink, blue, yellow). \textit{Left/Right panels:} Linear or log scale are adopted to better highlight the central and envelope regions, respectively. The location of the border of the overshooting region is indicated by vertical lines (dotted-dashed and dotted for hydro and magnetic models, respectively).  \label{fig:density_radius_Z}}
 \end{figure*}

The metallicity is changed within its observational error bars: values of 0.0156 and 0.0215 are tested. We also computed models adopting as $Y$ the solar value +0.05 and +0.10. The impact of $Z$ and $Y$ on stellar evolution is shown in Fig.\,\ref{fig:zhe} and summarised in Table\,\ref{table:influence} while the density profile and density function are shown in Fig.\,\ref{fig:density_radius_Z} for models at $R=R_\star$. 

An increase in $Z$ leads to a slight increase in $\dot{M}$ as well as an increase in $N/H$. The larger $N/H$ at $R_\star$ is explained by the initial larger $N/H$. Consequently, $k_2$ is slightly lower at larger $Z$ despite the model being younger and having evolved less than its lower-$Z$ counterpart. Models with larger $Z$ evolve at lower $T_\text{eff}$ and $L$. None of the modified-$Z$ models are able to reproduce all stellar properties simultaneously. 
 
An increase in $Y$ leads to a significant increase in $\dot{M}$ due to the larger $T_\text{eff}$ and $L$. Yet, the mass at $R_\star$ is not significantly impacted by the larger $\dot{M}$ owing to the fact that models are also much younger than their $Y_\odot$ counterparts. Models with larger $Y$ also show larger $N/H$; This effect is more pronounced in the magnetic models than in the hydro models. Models at higher $Y$ evolve faster to reach $R_\star$ and bluer in the Hertzsprung-Russell diagram and their internal density profile is significantly more pronounced than models at $Y_\odot$, hence their lower $k_2$. While increasing the initial helium content $Y$ goes in the right direction to solve the $k_2$-discrepancy, very large values -- not observationally supported -- would be necessary to reproduce $k_{2,\star}$.

\subsection{Initial mass \label{subsect:initial_mass}}
We built tracks with $M_\text{ini}$\,$\sim$\,$29.8$ and $\sim$\,$30.6\,M_\odot$ so that they have $M(R_\star)$\,$\sim$\,$29.1$ and $\sim$\,$30.0\,M_\odot$, corresponding to the lower and upper limits on $M_\star$. The impact on the evolution of parameters is summarised in Table\,\ref{table:influence} and shown in Fig.\,\ref{fig:initial_mass}. Models with the lower $M_\text{ini}$ predict $\pm$\,$200$\,K (0.4\%) lower effective temperatures and $\pm$\,$10000\,L_\odot$ (6\%) lower bolometric luminosity. The mass-loss rate is $10^{-7}\,M_\odot$\,yr$^{-1}$ (5\%) lower. The impact on $N/H$ is not significant. The predicted age is 0.1\,Myr (2\%) larger as the initially less massive star needs to evolve longer than the reference star to reach $R_\star$. Consequently, $k_2$ is slightly, but not significantly, lower as the star has evolved longer. Initially less massive models show a larger central density and the hook in the density profile happens at a lower $r/R$; This effect goes in the same direction as the effect of the age. Models with the larger $M_\text{ini}$ show the exact opposite quantitative behaviour.

\subsection{Mass-loss rate $\dot{M}$ \label{subsect:masslossrate}}
If clumping starts close to the photosphere, theoretical prescriptions underestimate mass-loss rates by a factor $\xi=C_\text{c}^{0.25}$, with $C_\text{c}=8-10$. We computed binary-star models adopting the mass-loss rate of \citet{krticka24} multiplied by $\xi=1.78$ (see Sect.\,\ref{sect:models}). The evolution of parameters are shown in Fig.\,\ref{fig:zhe} and summarised in Table\,\ref{table:influence} while the density contrast and density function of the model at $R=R_\star$ are shown in Fig.\,\ref{fig:density_radius_Z}. Because $\dot{M}$ is larger, models have a slightly larger initial mass of $32.0\,M_\odot$. These models thus evolve at higher $T_\text{eff}$ and $L$ in the Hertzsprung-Russell diagram and reach $R_\star$ at a younger age. The impact on $k_2$ is negligible. Indeed, the star is initially more massive so its $k_2$ decreases faster, but it reaches $R_\star$ earlier thus has not had time to evolve as much as a less massive star. Its density contrast is thus not much different from the less massive star at $R = R_\star$. Consequently, even if the mass-loss rate is underestimated due to clumping, it has no significant impact on our results. It demonstrates the robustness of  $k_2$ as indicator of stellar interiors as it is not sensitive to the mass-loss rate prescription we adopt (to a certain limit of reasonable mass-loss rates). We also tested models with the same mass-loss rate prescription but initially slightly more or less massive in such a way that their mass at $R_\star$ is still equal to $M_\star$ within the error bars. The impact on the results models is negligible, in line with the results of Sect.\,\ref{subsect:initial_mass}.

\subsection{Mixing length parameter $\alpha_\text{MLT}$ \label{subsect:alpha_MLT}}
\begin{figure}
\centering
\includegraphics[width=0.75\linewidth]{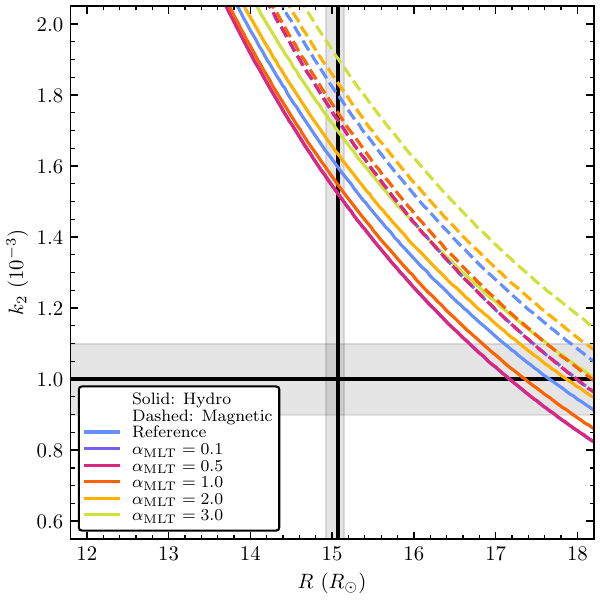}
\caption{Evolution of $k_2$ with stellar radius for hydro (plain lines) and magnetic (dashed lines) binary-star models of $M_\text{ini} = 30.2M_\odot$, $\alpha_\text{ov}=0.5$, $Z=0.0183$, $Y_\odot$, and \citet{krticka24} mass-loss rate for different values of $\alpha_\text{MLT}$. Observational values and their error bars are shown (plain black line and grey shaded area).  \label{fig:amlt}}
\end{figure}

We built models adopting different values for the mixing length parameter $\alpha_\text{MLT}$, namely 0.1, 0.5, 2.0, and 3.0. In main-sequence massive stars, $\alpha_\text{MLT}$ only impacts the small helium and iron convective zones in the envelope. While the value of $\alpha_\text{MLT}$ has no significant impact on $M, T_\text{eff}, L_\text{bol}, V_\text{surf}, N/H, \dot{M}$, and the age (not shown), it has a non-negligible impact on $k_2$ (see Fig.\,\ref{fig:amlt}). The lower the $\alpha_\text{MLT}$, the lower the $k_2$. The evolution of $k_2$ with $\alpha_\text{MLT} = 0.1$ and 0.5 overlap. Neither the density profile nor density function in the envelope show a monotonic behaviour with $\alpha_\text{MLT}$. Generally, differences are small as $\alpha_\text{MLT}$ only affects tiny regions. Thus, even a large change in $\alpha_\text{MLT}$ only has a limited impact on $k_2$. We conclude that even if the $\alpha_\text{MLT}$ value was revised upwards or downwards, it would not have any important consequence on our models and conclusions.

\subsection{Best-fit models}
\begin{figure}
\includegraphics[width=\linewidth]{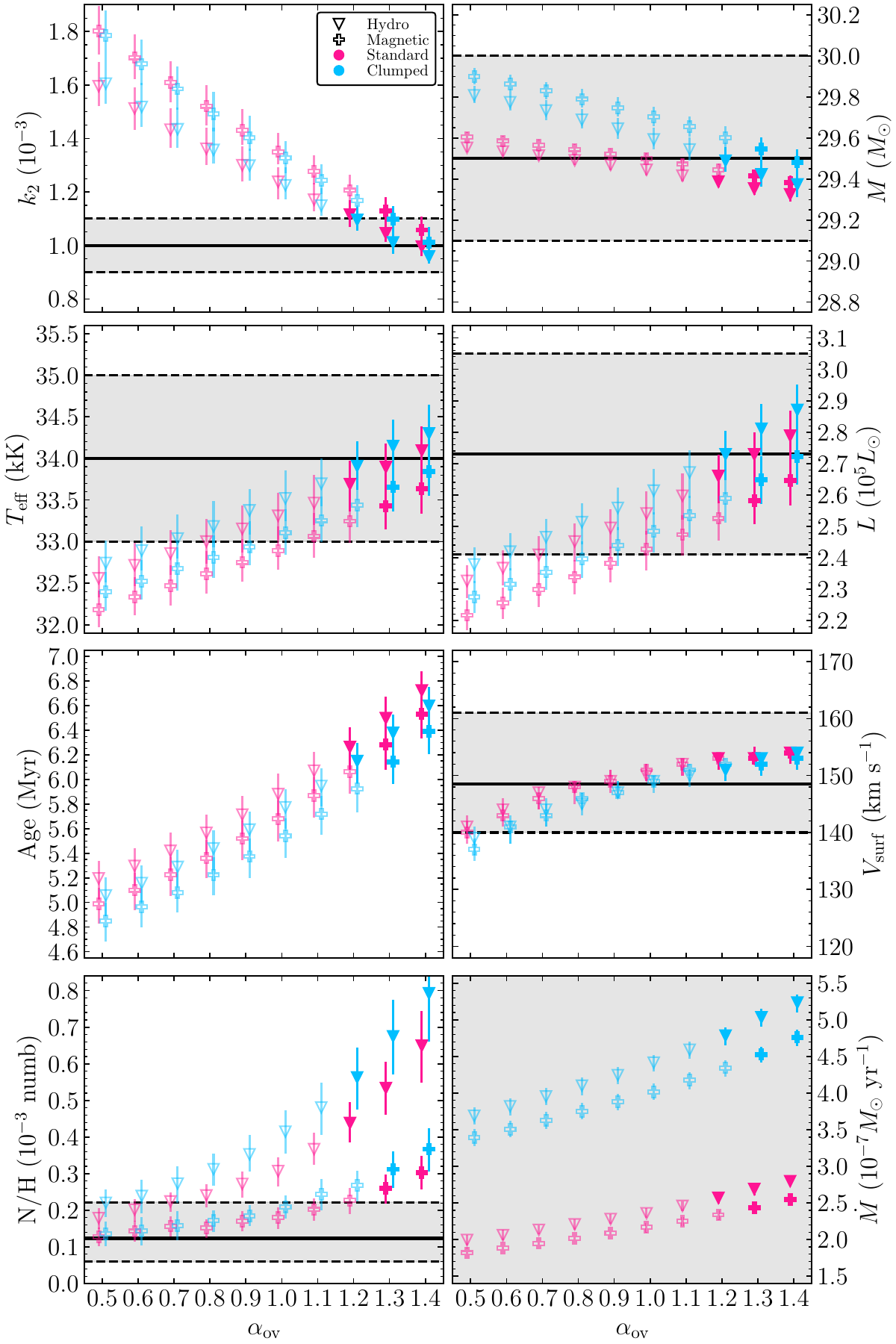}
\caption{Best-fit models in terms of radius: Stellar parameters as a function of $\alpha_\text{ov}$. Models that fit $k_{2,\star}, M_\star, T_{\text{eff},\star}, L_{\text{bol},\star}, V_{\text{surf},\star}$, and $\dot{M}_\star$ are shown with filled symbols. Error bars include the error bars on $R_\star$ and $Z_\star$ (the latter dominates over the former). Observational values and their error bars are shown (plain and dashed black lines). \label{fig:best-fit}}
\end{figure}

Best-fit standard and clumped models are shown in Fig.\,\ref{fig:best-fit} as a function of $\alpha_\text{ov}$. The error bars on the best-fit parameters take into account both the error bars on $R_\star$ and $Z_\star$: We consider when the models at $Z_\star, Z_\star + \delta Z_\star$, and $Z_\star - \delta Z_\star$ reproduce $R_\star \pm \delta R_\star$. The differences in the models obtained when adopting different $M_\text{ini}$ are not critical to assess which models reproduce the observations. Therefore, we do not consider them in the error bars of the best-fit models, as considering them would require to build a dedicated grid of models beyond the scope of this paper. Both standard and clumped hydro models reproduce all properties of the stars but $N/H$ for $\alpha_\text{ov} \ge 1.2$ while the magnetic counterparts necessitate $\alpha_\text{ov} \ge 1.3$. At these large $\alpha_\text{ov}$, magnetic models predict $N/H$ close to the observed values, while hydro models predict values about twice as large.

\section{\texttt{PoWR} atmosphere models\label{sect:powr}}
We demonstrated in previous sections the importance of predicting the correct stellar radius. We investigated whether the approximate treatment of the atmosphere in \texttt{GENEC} can hamper the determination of the radius. We supplemented the structures of ten well-chosen models on the reference \texttt{GENEC} binary-star track ($\alpha_\text{ov}=0.5$, $Z=0.0183$, $Y_\odot$, and \citet{krticka24} mass-loss rate) with detailed atmosphere models using the Postdam Wolf-Rayet code \citep[\texttt{PoWR},][]{grafener02, sander15}. We refer to \citet{josiek25} for a detailed explanation of the physics of the models, specifically their so-called deep models. The \texttt{PoWR} models capture the physics of radially-expanding media in non-local thermodynamic equilibrium. In these models, the atmosphere is structurally split into a hydrostatic domain and a wind domain with a prescribed velocity $\beta$-law (with $\beta=1$). We chose $\log\rho=-7.0$ as the lower boundary condition on the density of the models so that the boundary is located below the iron bump. In all ten models, the atmosphere is optically thin and does not produce any significant wind lines in the optical, which is consistent with observations \citep{rosu20b}. Both density and opacity profiles from the \texttt{PoWR} models agree with the \texttt{GENEC} models in the domain covered by the latter (see Fig.\,\ref{fig:powr_GENEC}), meaning that the atmospheric parameters predicted by \texttt{GENEC} (e.g., $T_\text{eff}$) are consistent with the more detailed \texttt{PoWR} atmosphere models. The expansion of the stellar envelope is driven from the interior, not from the surface, as shown by the lack of any prominent surface-inflated region that could possibly be interpreted as a thick wind (see Fig.\,\ref{fig:powr}), as opposed to structures of very massive stars \citep[see figure 3 in][]{josiek25}. As is typical for O stars, the \texttt{PoWR} model places the iron-bump region below the wind, that is to say, in the hydrostatic layers below the photosphere, where there is no strong departure from local thermodynamic equilibrium. This means that the Rosseland mean opacity from OPAL used by \texttt{GENEC} to compute the hydrostatic structure is a good approximation for the flux-weighted mean opacity (which is directly proportional to the radiative acceleration) calculated by \texttt{PoWR} in these layers.


\section{Discussion: possible misalignment of the stellar rotation axes\label{sect:misalignment}}
The difficulty to reconcile the theoretical and observational $k_2$ of the stars may suggest a bias in our interpretation of the apsidal motion in terms of $k_2$. A potential cause could be a misalignment between the stellar rotation axes and the normal to the orbital plane. In that case, the Newtonian contribution to the apsidal motion rate takes the general expression given by Eq.\,(3) of \citet{shakura85}, simplified for the case of a twin system for which we assume the stars have their rotation axes mutually parallel: 
\begin{equation}
\dot\omega_\text{N}=\frac{4\pi}{P_\text{orb}} k_{2,\text{mis}}\left(\frac{R_\star}{a}\right)^5 \left[15f-\frac{F_\alpha}{\sin^2i}2g\left(\frac{P_\text{orb}}{P_{\text{rot},\star}}\right)^2\right],
\label{eqn:omegadot_misaligned}
\end{equation}
where $-\frac{F_\alpha}{\sin^2 i}$ is introduced compared to Eq.\,\eqref{eqn:omegadot}, with
\begin{equation}
F_\alpha = \cos\alpha(\cos\alpha-\cos\beta\cos i) + \frac{1}{2} \sin^2 i \left(1-5\cos^2\alpha\right),
\label{eqn:falpha}
\end{equation}
where $\alpha$ is the angle between the star rotation axis and the normal to the orbital plane and $\beta$ is the angle between the star rotation axis and the line of sight to the binary centre. 

We used the cosine relationship in spherical trigonometry to express $\cos\beta$ in terms of $\alpha, i$, and the azimutal angle of the rotation axes of the stars $\theta$:
\begin{equation}
\cos\beta = \cos i \cos\alpha + \sin i \sin\alpha \cos\theta.
\end{equation}
When $\theta = 0^\circ$ (thus $\beta=\alpha+i$) and $\theta=180^\circ$ (thus $\beta=\alpha-i$), the rotation axes, the normal to the orbital plane, and the line of sight are coplanar and $F_\alpha$ takes its extremum values \citep{rosu22a}. In these extreme cases, the term in brackets in Eq.\,\eqref{eqn:omegadot_misaligned} becomes minimal and $k_{2,\text{mis}}$ has to be larger to compensate. 

\begin{figure}
\centering
\includegraphics[width=\linewidth]{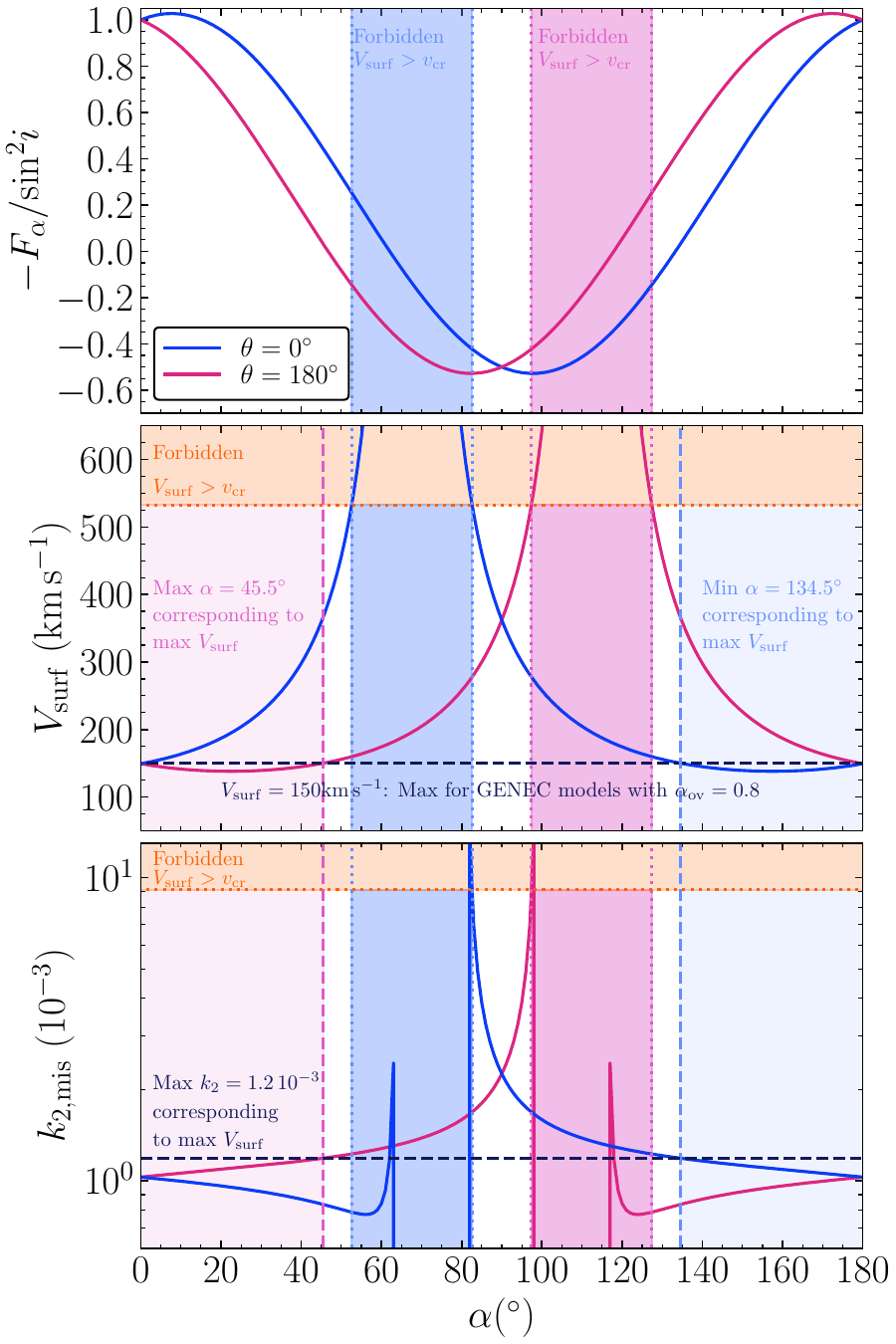}
\caption{Dependence of $-\frac{F_\alpha}{\sin^2 i}$ (from Eq.\,\eqref{eqn:falpha}, \textit{top panel}), $V_\text{surf}$ (\textit{middle panel}), and $k_{2,\text{mis}}$ (from Eq.\,\eqref{eqn:omegadot_misaligned}, \textit{bottom panel}) with $\alpha$ for the specific case of $\theta=0^\circ$ (blue) and $\theta = 180^\circ$ (pink) and with $i=67.6^\circ$. The forbidden region of $V_\text{surf} > v_\text{cr}$ (in orange in the middle panel) translates into forbidden regions for $\alpha$ (in blue and pink delimited by dotted lines for $\theta=0^\circ$ and $180^\circ$, respectively). Consequently, $k_{2,\text{mis}} \le 0.9\,10^{-3}$ (forbidden region in orange in the bottom panel). The pseudo-synchronisation constraint applied to \texttt{GENEC} hydro binary-star models with $\alpha_\text{ov} = 0.8$ imposes that $V_\text{surf} \le 150$\,km\,s$^{-1}$ (navy blue dashed line in the middle panel). This constraint translates into a constraint on $\alpha$: $\alpha \ge 134.5^\circ$ if $\theta=0^\circ$ and $\alpha \le 45.5^\circ$ if $\theta=180^\circ$ (light blue and light pink area, respectively). This sets a maximum value for $k_{2,\text{mis}} = 1.2\,10^{-3}$ (navy blue horizontal dashed line in the bottom panel). \label{fig:falpha_k2}}
\end{figure}

We show in Fig.\,\ref{fig:falpha_k2} (top panel) the dependence of $-\frac{F_\alpha}{\sin^2 i}$ with $\alpha$ for the specific cases of $\theta=0^\circ$ and $180^\circ$ and with $i=67.6^\circ$ for the system \citep{rosu20b}: $F_\alpha$ is minimum when $\alpha = 97^\circ$ and $83^\circ$, respectively, and maximum when $\alpha = 8^\circ$ and $172^\circ$, respectively. 

We also show in Fig.\,\ref{fig:falpha_k2} (middle panel) the dependence of $V_\text{surf} = \frac{v_\text{eq}\sin\beta}{\sin\beta}$ with $\alpha$. The condition of subcritical stellar rotation restricts the possible values of $\alpha$. The condition writes 
\begin{equation}
\Omega_\text{surf} \le \Omega_\text{cr} = \sqrt{\frac{GM_\star}{R_\star^3}(1-\Gamma_e)},
\label{eqn:omega}
\end{equation}
with 
\begin{equation}
\Gamma_e = \frac{\kappa_e L_{\text{bol},\star}}{4\pi cGM_\star}
\end{equation}
the Eddington factor, and $\kappa_e = 0.34$\,cm$^2$\,g$^{-1}$ the electron scattering opacity \citep{sanyal15}. Eq.\,\eqref{eqn:omega} further translates into a constraint on the equatorial velocity 
\begin{equation}
v_\text{eq} \le 436.7 \sqrt{\frac{M_\star R_\odot}{M_\odot R_\star} \left(1-2.6\,10^{-5}\frac{L_{\text{bol},\star} M_\odot}{L_\odot M_\star}\right)} = 532\, \text{km}\,\text{s}^{-1}.
\end{equation}
From the projected rotational velocities $v_{\text{eq}}\sin\beta = 138$\,km\,s$^{-1}$ \citep{rosu22b}, we obtain the constraint that $\beta$\,$\ge$\,$15^\circ$ and $\le$\,$165^\circ$.  For the two extreme cases of $\theta$\,$=$\,$0^\circ$ and $180^\circ$, this translates into the constraint that $\alpha$\,$\le$\,$53^\circ$ and $\ge$\,$83^\circ$, and $\alpha$\,$\le$\,$97^\circ$ and $\ge$\,$127^\circ$, respectively, as illustrated in Fig.\,\ref{fig:falpha_k2} (the blue and pink region framed by dotted lines show the forbidden regions of $\alpha$).

We computed $k_{2,\text{mis}}$ for values of $\alpha$ ranging from $0^\circ$ to $180^\circ$. As illustrated in Fig.\,\ref{fig:falpha_k2} (bottom panel), if we restrict to the allowed regions, $k_{2,\text{mis}}$ takes its maximum value of $9.1\,10^{-3}$ and minimum value of $0.8\,10^{-3}$ when $\alpha$\,$=$\,$83^\circ$ and $53^\circ$ when $\theta$\,$=$\,$0^\circ$, and when $\alpha$\,$=$\,$97^\circ$ and $127^\circ$ when $\theta$\,$=$\,$180^\circ$. 

The reference hydro binary-star model with $\alpha_\text{ov}$\,$=$\,$0.5$ predicts $k_2$\,$=$\,$1.6\,10^{-3}$. If it was to reproduce $k_{2,\text{mis}}$, the star rotation axes should be misaligned by an angle $\alpha$\,$\in$\,$[83^\circ,100^\circ]$ if $\theta$\,$=$\,$0^\circ$ or $\alpha$\,$\in$\,$[80^\circ,97^\circ]$ if $\theta$\,$=$\,$180^\circ$. Such a high misalignment angle would be surprising in a short-period system. Indeed, \citet{marcussen22} showed that among the 43 binaries in their sample, only three show a large misalignment angle between the star rotation axes and the normal to the orbital plane. Yet, these systems have either larger orbital periods or lower total mass; Their tides are thus less efficient than in HD\,152248 to align their rotation axes with the orbital axis. Furthermore, such a high misalignment angle would give values of $V_\text{surf}$\,$\in$\,$[278, v_\text{crit}]$\,km\,s$^{-1}$ instead of the value $149$\,km\,s$^{-1}$  obtained in the case where $\beta=i$, as shown in Fig.\,\ref{fig:falpha_k2}. The system would be very far from pseudo-synchronisation, with a rotation period twice smaller than the orbital period. For these reasons, we conclude that the stars cannot be misaligned by an angle $\alpha$\,$\in$\,$[80^\circ, 100^\circ]$ and models with $\alpha_\text{ov}$\,$=$\,$0.5$ cannot reproduce $k_{2,\star}$. 

Given the efficiency of tides to synchronise systems \citep{zahn77, hut81} (see also Sect.\,\ref{subsect:not_synchronised}), we consider the constraint of pseudo-synchronisation as a good assumption for the system. For a value of $\alpha_\text{ov}$\,$=$\,$0.8$ suggested by \citet{scott21} for a $30\,M_\odot$-star, hydro binary-star models predict a value of $1.4\,10^{-3}$ for $k_2$, while the pseudo-synchronisation constraint imposes a value of $V_\text{surf}$\,$\sim$\,$150$\,km\,s$^{-1}$. This limit is shown in Fig.\,\ref{fig:falpha_k2} (horizontal dashed navy blue lines); It corresponds to a maximum misalignment angle $\alpha$\,$=$\,$45.5^\circ$ (light blue and pink regions in Fig.\,\ref{fig:falpha_k2}). For these misalignment angles, $k_{2,\text{mis}}$\,$=$\,$1.2\,10^{-3}$. This value is smaller than the value predicted by the model; Hence $\alpha_\text{ov}$\,$=$\,$0.8$ is still not sufficient. A hydro model requires a minimum $\alpha_\text{ov}$-value of 1.0, under the condition that $\alpha$\,$=$\,$ 47^\circ$, to reproduce $k_{2,\star}$.

We conclude that even if the axes were misaligned, it would not solve the $k_2$-discrepancy and a high $\alpha_\text{ov}$ of minimum 1.0 in the best case scenario would be necessary to reproduce the observations. Furthermore, given that both the synchronisation and alignment timescales are proportional to $(R/a)^6$ \citep{hut81, sciarini24}, we expect alignment if pseudo-synchronisation is achieved. Larger $\alpha_\text{ov}$ are thus necessary (Sect.\,\ref{subsect:alpha_ov}).

\section{Conclusion\label{sect:conclusion}}
We used the apsidal motion observed in the massive twin binary star HD\,152248 and its link to the internal stellar structure constants $k_2$ of the stars to probe their interiors. We aimed at constraining the internal mixing mechanisms and convective boundary mixing. 

We built stellar evolution models with the \texttt{GENEC} code assuming two different angular momentum transport schemes: purely hydrodynamic models (hydro) and magneto-diffusive models (magnetic). We showed that because tides very quickly synchronise the angular velocity of the star with its orbital motion, binary-star models suppress mixing inside stars compared to their single-star counterparts. We thus demonstrated that it is crucial to take into account binary effects through tides in stellar models as it changes the angular momentum evolution of the stars compared to a single-star counterpart. We investigated the impact of the overshooting parameter, metallicity, initial helium content, initial mass, mass-loss rate, and mixing length parameter on the evolution of stellar parameters. 

We showed that models computed with standard assumptions systematically predict larger $k_2$ than the observational value, meaning that models predict stars with too low a density contrast between their core and external layers, while observations suggest the core is more contracted and the envelope more extended. Our results suggest that both hydro and magnetic models have difficulties to reproduce the observed parameters. We demonstrated that reproducing $k_2$ and the apsidal motion rate simultaneously with the stellar parameters requires more mixing through larger overshooting -- $\alpha_\text{ov} =1.2$ and 1.3 for hydro and magnetic models, respectively --, while other parameters have a limited impact. The larger convective boundary mixing acts as increasing the main-sequence lifetime of the star and lowering the rate of expansion of the radius: The star thus has more time to contrast its density profile between the core and external layers. Even if the mass-loss rate was underestimated by a factor 1.78 due to clumping starting close to the photosphere, it would have no impact on the stellar parameters evolution, including $k_2$. 

We built dedicated atmosphere models with the Postdam Wolf-Rayet code (\texttt{PoWR}) to supplement the structures of the \texttt{GENEC} models. We showed that the atmospheric parameters predicted by \texttt{GENEC} are consistent with those from the detailed \texttt{PoWR} models, as both density and opacity profiles agree between the two codes. 

We discussed the impact of a possible misalignment of the stellar rotation axes with the normal to the orbital plane on our interpretation of the apsidal motion in terms of $k_2$. Given the efficiency of tides to synchronise short period systems as HD\,152248, we consider that the assumption of pseudo-synchronisation is sound for this system. It sets a constraint on the misalignment angle of stellar rotation axes of $45^\circ-52^\circ$ maximum, depending on the models. Even with such large angles -- unexpected as alignment happens on the same timescale as synchronisation --, it does not solve the $k_2$-discrepancy and a large overshooting parameter of minimum 1.0 is necessary to reproduce the observations.

In conclusion, this large overshooting indicates that the core-boundary mixing should be more important than suggested by standard models.  Whether it is an extension of the convective core or a strong mixing in the radiative zone close to the core is to be decided. One possibility could be mixing induced by gravity waves excited by tidal friction. 

\begin{acknowledgement}
The authors thank the anonymous referee for their constructive report. S.R. thanks Prof. C. Aerts for `The Beauty of $k_2$'. S.R., L.S., and S.E. acknowledges support from the Swiss National Science Foundation SNF project N$^\circ$212143. P.E. acknowledges support from the SNF project N$^\circ$219745. J.J. acknowledges funding support from the Deutsche Forschungsgemeinschaft (DFG, German Research Foundation) Project-ID 496854903 (SA4064/2-1, PI Sander).

\end{acknowledgement}


\bibliographystyle{aa}
\bibliography{extra}

\begin{thebibliography}{73}
\expandafter\ifx\csname natexlab\endcsname\relax\def\natexlab#1{#1}\fi

\bibitem[{{Abbott} {et~al.}(2016){Abbott}, {Abbott}, {Abbott}, {Abernathy},
  {Acernese}, {Ackley}, {Adams}, {Adams}, {Addesso}, {Adhikari}, {Adya},
  {Affeldt}, {Agathos}, {Agatsuma}, {Aggarwal}, {Aguiar}, {Aiello}, {Ain},
  {Ajith}, {Allen}, {Allocca}, {Altin}, {Anderson}, {Anderson}, {Arai},
  {Arain}, {Araya}, {Arceneaux}, {Areeda}, {Arnaud}, {Arun}, {Ascenzi},
  {Ashton}, {Ast}, {Aston}, {Astone}, {Aufmuth}, {Aulbert}, {Babak}, {Bacon},
  {Bader}, {Baker}, {Baldaccini}, {Ballardin}, {Ballmer}, {Barayoga},
  {Barclay}, {Barish}, {Barker}, {Barone}, {Barr}, {Barsotti}, {Barsuglia},
  {Barta}, {Bartlett}, {Barton}, {Bartos}, {Bassiri}, {Basti}, {Batch},
  {Baune}, {Bavigadda}, {Bazzan}, {Behnke}, {Bejger}, {Belczynski}, {Bell},
  {Bell}, {Berger}, {Bergman}, {Bergmann}, {Berry}, {Bersanetti}, {Bertolini},
  {Betzwieser}, {Bhagwat}, {Bhandare}, {Bilenko}, {Billingsley}, {Birch},
  {Birney}, {Birnholtz}, {Biscans}, {Bisht}, {Bitossi}, {Biwer}, {Bizouard},
  {Blackburn}, {Blair}, {Blair}, {Blair}, {Bloemen}, {Bock}, {Bodiya}, {Boer},
  {Bogaert}, {Bogan}, {Bohe}, {Bojtos}, {Bond}, {Bondu}, {Bonnand}, {Boom},
  {Bork}, {Boschi}, {Bose}, {Bouffanais}, {Bozzi}, {Bradaschia}, {Brady},
  {Braginsky}, {Branchesi}, {Brau}, {Briant}, {Brillet}, {Brinkmann},
  {Brisson}, {Brockill}, {Brooks}, {Brown}, {Brown}, {Brown}, {Buchanan},
  {Buikema}, {Bulik}, {Bulten}, {Buonanno}, {Buskulic}, {Buy}, {Byer},
  {Cabero}, {Cadonati}, {Cagnoli}, {Cahillane}, {Bustillo}, {Callister},
  {Calloni}, {Camp}, {Cannon}, {Cao}, {Capano}, {Capocasa}, {Carbognani},
  {Caride}, {Diaz}, {Casentini}, {Caudill}, {Cavagli{\`a}}, {Cavalier},
  {Cavalieri}, {Cella}, {Cepeda}, {Baiardi}, {Cerretani}, {Cesarini},
  {Chakraborty}, {Chalermsongsak}, {Chamberlin}, {Chan}, {Chao}, {Charlton},
  {Chassande-Mottin}, {Chen}, {Chen}, {Cheng}, {Chincarini}, {Chiummo}, {Cho},
  {Cho}, {Chow}, {Christensen}, {Chu}, {Chua}, {Chung}, {Ciani}, {Clara},
  {Clark}, {Cleva}, {Coccia}, {Cohadon}, {Colla}, {Collette}, {Cominsky},
  {Constancio}, {Conte}, {Conti}, {Cook}, {Corbitt}, {Cornish}, {Corsi},
  {Cortese}, {Costa}, {Coughlin}, {Coughlin}, {Coulon}, {Countryman},
  {Couvares}, {Cowan}, {Coward}, \& {Cowart}}]{abbott16}
{Abbott}, B.~P., {Abbott}, R., {Abbott}, T.~D., {et~al.} 2016, \prl, 116,
  061102

\bibitem[{{Aerts}(2021)}]{aerts21}
{Aerts}, C. 2021, Reviews of Modern Physics, 93, 015001

\bibitem[{{Asplund} {et~al.}(2009){Asplund}, {Grevesse}, {Sauval}, \&
  {Scott}}]{asplund09}
{Asplund}, M., {Grevesse}, N., {Sauval}, A.~J., \& {Scott}, P. 2009, \araa, 47,
  481

\bibitem[{{Baraffe} {et~al.}(2023){Baraffe}, {Clarke}, {Morison}, {Vlaykov},
  {Constantino}, {Goffrey}, {Guillet}, {Le Saux}, \& {Pratt}}]{baraffe23}
{Baraffe}, I., {Clarke}, J., {Morison}, A., {et~al.} 2023, \mnras, 519, 5333

\bibitem[{{Bj{\"o}rklund} {et~al.}(2021){Bj{\"o}rklund}, {Sundqvist}, {Puls},
  \& {Najarro}}]{bjorklund21}
{Bj{\"o}rklund}, R., {Sundqvist}, J.~O., {Puls}, J., \& {Najarro}, F. 2021,
  \aap, 648, A36

\bibitem[{{Brands} {et~al.}(2022){Brands}, {de Koter}, {Bestenlehner},
  {Crowther}, {Sundqvist}, {Puls}, {Caballero-Nieves}, {Abdul-Masih},
  {Driessen}, {Garc{\'\i}a}, {Geen}, {Gr{\"a}fener}, {Hawcroft}, {Kaper},
  {Keszthelyi}, {Langer}, {Sana}, {Schneider}, {Shenar}, \& {Vink}}]{brands22}
{Brands}, S.~A., {de Koter}, A., {Bestenlehner}, J.~M., {et~al.} 2022, \aap,
  663, A36

\bibitem[{{Brott} {et~al.}(2011){Brott}, {de Mink}, {Cantiello}, {Langer}, {de
  Koter}, {Evans}, {Hunter}, {Trundle}, \& {Vink}}]{brott11}
{Brott}, I., {de Mink}, S.~E., {Cantiello}, M., {et~al.} 2011, \aap, 530, A115

\bibitem[{{Castro} {et~al.}(2014){Castro}, {Fossati}, {Langer},
  {Sim{\'o}n-D{\'\i}az}, {Schneider}, \& {Izzard}}]{castro14}
{Castro}, N., {Fossati}, L., {Langer}, N., {et~al.} 2014, \aap, 570, L13

\bibitem[{{Chaboyer} \& {Zahn}(1992)}]{chaboyer92}
{Chaboyer}, B. \& {Zahn}, J.-P. 1992, \aap, 253, 173

\bibitem[{{Chieffi} \& {Limongi}(2020)}]{chieffi20}
{Chieffi}, A. \& {Limongi}, M. 2020, \apj, 890, 43

\bibitem[{{Claret} \& {Torres}(2019)}]{claret19}
{Claret}, A. \& {Torres}, G. 2019, \apj, 876, 134

\bibitem[{{Dias} {et~al.}(2021){Dias}, {Monteiro}, {Moitinho}, {L{\'e}pine},
  {Carraro}, {Paunzen}, {Alessi}, \& {Villela}}]{dias21}
{Dias}, W.~S., {Monteiro}, H., {Moitinho}, A., {et~al.} 2021, \mnras, 504, 356

\bibitem[{{Eddington}(1925)}]{eddington25}
{Eddington}, A.~S. 1925, The Observatory, 48, 73

\bibitem[{{Eggenberger} {et~al.}(2008){Eggenberger}, {Meynet}, {Maeder},
  {Hirschi}, {Charbonnel}, {Talon}, \& {Ekstr{\"o}m}}]{eggenberger08}
{Eggenberger}, P., {Meynet}, G., {Maeder}, A., {et~al.} 2008, Ap\&SS, 316, 43

\bibitem[{{Eggenberger} {et~al.}(2022){Eggenberger}, {Moyano}, \& {den
  Hartogh}}]{eggenberger22}
{Eggenberger}, P., {Moyano}, F.~D., \& {den Hartogh}, J.~W. 2022, \aap, 664,
  L16

\bibitem[{{Ekstr{\"o}m} {et~al.}(2012){Ekstr{\"o}m}, {Georgy}, {Eggenberger},
  {Meynet}, {Mowlavi}, {Wyttenbach}, {Granada}, {Decressin}, {Hirschi},
  {Frischknecht}, {Charbonnel}, \& {Maeder}}]{ekstrom12}
{Ekstr{\"o}m}, S., {Georgy}, C., {Eggenberger}, P., {et~al.} 2012, \aap, 537,
  A146

\bibitem[{{Ertl} {et~al.}(2016){Ertl}, {Janka}, {Woosley}, {Sukhbold}, \&
  {Ugliano}}]{ertl16}
{Ertl}, T., {Janka}, H.-T., {Woosley}, S.~E., {Sukhbold}, T., \& {Ugliano}, M.
  2016, \apj, 818, 124

\bibitem[{{Fellay} {et~al.}(2024){Fellay}, {Dupret}, \& {Rosu}}]{fellay24}
{Fellay}, L., {Dupret}, M.~A., \& {Rosu}, S. 2024, A\&A, 683, A210

\bibitem[{{Fragos} {et~al.}(2023){Fragos}, {Andrews}, {Bavera}, {Berry},
  {Coughlin}, {Dotter}, {Giri}, {Kalogera}, {Katsaggelos}, {Kovlakas},
  {Lalvani}, {Misra}, {Srivastava}, {Qin}, {Rocha}, {Rom{\'a}n-Garza}, {Serra},
  {Stahle}, {Sun}, {Teng}, {Trajcevski}, {Tran}, {Xing}, {Zapartas}, \&
  {Zevin}}]{fragos23}
{Fragos}, T., {Andrews}, J.~J., {Bavera}, S.~S., {et~al.} 2023, \apjs, 264, 45

\bibitem[{{Gilkis} {et~al.}(2021){Gilkis}, {Shenar}, {Ramachandran}, {Jermyn},
  {Mahy}, {Oskinova}, {Arcavi}, \& {Sana}}]{gilkis21}
{Gilkis}, A., {Shenar}, T., {Ramachandran}, V., {et~al.} 2021, \mnras, 503,
  1884

\bibitem[{{Gr{\"a}fener} {et~al.}(2002){Gr{\"a}fener}, {Koesterke}, \&
  {Hamann}}]{grafener02}
{Gr{\"a}fener}, G., {Koesterke}, L., \& {Hamann}, W.-R. 2002, \aap, 387, 244

\bibitem[{{Hawcroft} {et~al.}(2024){Hawcroft}, {Mahy}, {Sana}, {Sundqvist},
  {Abdul-Masih}, {Brands}, {Decin}, {de Koter}, \& {Puls}}]{hawcroft24}
{Hawcroft}, C., {Mahy}, L., {Sana}, H., {et~al.} 2024, \aap, 690, A126

\bibitem[{{Hejlesen}(1987)}]{hejlesen87}
{Hejlesen}, P.~M. 1987, A\&AS, 69, 251

\bibitem[{{Higgins} \& {Vink}(2019)}]{higgins19}
{Higgins}, E.~R. \& {Vink}, J.~S. 2019, \aap, 622, A50

\bibitem[{{Hut}(1981)}]{hut81}
{Hut}, P. 1981, \aap, 99, 126

\bibitem[{{Josiek} {et~al.}(2025){Josiek}, {Sander}, {Bernini-Peron},
  {Ekstr{\"o}m}, {Gonz{\'a}lez-Tor{\`a}}, {Lefever}, {Moens}, {Ramachandran},
  \& {Sch{\"o}sser}}]{josiek25}
{Josiek}, J., {Sander}, A.~A.~C., {Bernini-Peron}, M., {et~al.} 2025, \aap,
  697, A193

\bibitem[{{Joyce} \& {Tayar}(2023)}]{joyce23}
{Joyce}, M. \& {Tayar}, J. 2023, Galaxies, 11, 75

\bibitem[{{Kaiser} {et~al.}(2020){Kaiser}, {Hirschi}, {Arnett}, {Georgy},
  {Scott}, \& {Cristini}}]{kaiser20}
{Kaiser}, E.~A., {Hirschi}, R., {Arnett}, W.~D., {et~al.} 2020, \mnras, 496,
  1967

\bibitem[{{Krti{\v{c}}ka} \& {Kub{\'a}t}(2017)}]{krticka17}
{Krti{\v{c}}ka}, J. \& {Kub{\'a}t}, J. 2017, \aap, 606, A31

\bibitem[{{Krti{\v{c}}ka} \& {Kub{\'a}t}(2018)}]{krticka18}
{Krti{\v{c}}ka}, J. \& {Kub{\'a}t}, J. 2018, \aap, 612, A20

\bibitem[{{Krti{\v{c}}ka} {et~al.}(2024){Krti{\v{c}}ka}, {Kub{\'a}t}, \&
  {Krti{\v{c}}kov{\'a}}}]{krticka24}
{Krti{\v{c}}ka}, J., {Kub{\'a}t}, J., \& {Krti{\v{c}}kov{\'a}}, I. 2024, \aap,
  681, A29

\bibitem[{{Kuhn} {et~al.}(2017{\natexlab{a}}){Kuhn}, {Getman}, {Feigelson},
  {Sills}, {Gromadzki}, {Medina}, {Borissova}, \& {Kurtev}}]{kuhn17a}
{Kuhn}, M.~A., {Getman}, K.~V., {Feigelson}, E.~D., {et~al.}
  2017{\natexlab{a}}, \aj

\bibitem[{{Kuhn} {et~al.}(2017{\natexlab{b}}){Kuhn}, {Medina}, {Getman},
  {Feigelson}, {Gromadzki}, {Borissova}, \& {Kurtev}}]{kuhn17b}
{Kuhn}, M.~A., {Medina}, N., {Getman}, K.~V., {et~al.} 2017{\natexlab{b}}, \aj

\bibitem[{{Lennon} \& {Dufton}(1983)}]{lennon83}
{Lennon}, D.~J. \& {Dufton}, P.~L. 1983, \mnras, 203, 443

\bibitem[{{Limongi} \& {Chieffi}(2006)}]{limongi06}
{Limongi}, M. \& {Chieffi}, A. 2006, \apj, 647, 483

\bibitem[{{Maeder} \& {Meynet}(2003)}]{maeder03}
{Maeder}, A. \& {Meynet}, G. 2003, \aap, 411, 543

\bibitem[{{Maeder} \& {Meynet}(2004)}]{maeder04}
{Maeder}, A. \& {Meynet}, G. 2004, \aap, 422, 225

\bibitem[{{Maeder} \& {Meynet}(2005)}]{maeder05}
{Maeder}, A. \& {Meynet}, G. 2005, \aap, 440, 1041

\bibitem[{{Maeder} \& {Zahn}(1998)}]{maeder98}
{Maeder}, A. \& {Zahn}, J.-P. 1998, \aap, 334, 1000

\bibitem[{{Mao} {et~al.}(2024){Mao}, {Woodward}, {Herwig}, {Denissenkov},
  {Blouin}, {Thompson}, \& {McDermott}}]{mao24}
{Mao}, H., {Woodward}, P., {Herwig}, F., {et~al.} 2024, \apj, 975, 271

\bibitem[{{Marcussen} \& {Albrecht}(2022)}]{marcussen22}
{Marcussen}, M.~L. \& {Albrecht}, S.~H. 2022, ApJ, 933, 227

\bibitem[{{Martinet} {et~al.}(2021){Martinet}, {Meynet}, {Ekstr{\"o}m},
  {Sim{\'o}n-D{\'\i}az}, {Holgado}, {Castro}, {Georgy}, {Eggenberger},
  {Buldgen}, {Salmon}, {Hirschi}, {Groh}, {Farrell}, \& {Murphy}}]{martinet21}
{Martinet}, S., {Meynet}, G., {Ekstr{\"o}m}, S., {et~al.} 2021, \aap, 648, A126

\bibitem[{{Martins} \& {Palacios}(2013)}]{martins13}
{Martins}, F. \& {Palacios}, A. 2013, \aap, 560, A16

\bibitem[{{Mathys} {et~al.}(2002){Mathys}, {Andrievsky}, {Barbuy}, {Cunha}, \&
  {Korotin}}]{mathys02}
{Mathys}, G., {Andrievsky}, S.~M., {Barbuy}, B., {Cunha}, K., \& {Korotin},
  S.~A. 2002, \aap, 387, 890

\bibitem[{{Nandal} {et~al.}(2024){Nandal}, {Meynet}, {Ekstr{\"o}m}, {Moyano},
  {Eggenberger}, {Choplin}, {Georgy}, {Farrell}, \& {Maeder}}]{nandal24}
{Nandal}, D., {Meynet}, G., {Ekstr{\"o}m}, S., {et~al.} 2024, \aap, 684, A169

\bibitem[{{Paunzen} {et~al.}(2010){Paunzen}, {Heiter}, {Netopil}, \&
  {Soubiran}}]{paunzen10}
{Paunzen}, E., {Heiter}, U., {Netopil}, M., \& {Soubiran}, C. 2010, \aap, 517,
  A32

\bibitem[{{Pedersen} {et~al.}(2021){Pedersen}, {Aerts}, {P{\'a}pics},
  {Michielsen}, {Gebruers}, {Rogers}, {Molenberghs}, {Burssens}, {Garcia}, \&
  {Bowman}}]{pedersen21}
{Pedersen}, M.~G., {Aerts}, C., {P{\'a}pics}, P.~I., {et~al.} 2021, Nature
  Astronomy, 5, 715

\bibitem[{{Qin} {et~al.}(2018){Qin}, {Fragos}, {Meynet}, {Andrews},
  {S{\o}rensen}, \& {Song}}]{qin18}
{Qin}, Y., {Fragos}, T., {Meynet}, G., {et~al.} 2018, \aap, 616, A28

\bibitem[{{Rauw} {et~al.}(2016){Rauw}, {Rosu}, {Noels}, {Mahy}, {Schmitt},
  {Godart}, {Dupret}, \& {Gosset}}]{rauw16}
{Rauw}, G., {Rosu}, S., {Noels}, A., {et~al.} 2016, A\&A, 594, A33

\bibitem[{{Rosu} {et~al.}(2020a){Rosu}, {Noels}, {Dupret}, {Rauw}, {Farnir}, \&
  {Ekstr{\"o}m}}]{rosu20a}
{Rosu}, S., {Noels}, A., {Dupret}, M.~A., {et~al.} 2020a, A\&A, 642, A221

\bibitem[{{Rosu} {et~al.}(2020b){Rosu}, {Rauw}, {Conroy}, {Gosset}, {Manfroid},
  \& {Royer}}]{rosu20b}
{Rosu}, S., {Rauw}, G., {Conroy}, K.~E., {et~al.} 2020b, A\&A, 635, A145

\bibitem[{{Rosu} {et~al.}(2022a){Rosu}, {Rauw}, {Farnir}, {Dupret}, \&
  {Noels}}]{rosu22a}
{Rosu}, S., {Rauw}, G., {Farnir}, M., {Dupret}, M.~A., \& {Noels}, A. 2022a,
  A\&A, 660, A120

\bibitem[{{Rosu} {et~al.}(2022b){Rosu}, {Rauw}, {Naz{\'e}}, {Gosset}, \&
  {Sterken}}]{rosu22b}
{Rosu}, S., {Rauw}, G., {Naz{\'e}}, Y., {Gosset}, E., \& {Sterken}, C. 2022b,
  A\&A, 664, A98

\bibitem[{{Sana} {et~al.}(2007){Sana}, {Rauw}, {Sung}, {Gosset}, \&
  {Vreux}}]{sana07}
{Sana}, H., {Rauw}, G., {Sung}, H., {Gosset}, E., \& {Vreux}, J.~M. 2007,
  \mnras, 377, 945

\bibitem[{{Sander} {et~al.}(2015){Sander}, {Shenar}, {Hainich},
  {G{\'\i}menez-Garc{\'\i}a}, {Todt}, \& {Hamann}}]{sander15}
{Sander}, A., {Shenar}, T., {Hainich}, R., {et~al.} 2015, \aap, 577, A13

\bibitem[{{Sanyal} {et~al.}(2015){Sanyal}, {Grassitelli}, {Langer}, \&
  {Bestenlehner}}]{sanyal15}
{Sanyal}, D., {Grassitelli}, L., {Langer}, N., \& {Bestenlehner}, J.~M. 2015,
  \aap, 580, A20

\bibitem[{{Sciarini} {et~al.}(2024){Sciarini}, {Ekstr{\"o}m}, {Eggenberger},
  {Meynet}, {Fragos}, \& {Song}}]{sciarini24}
{Sciarini}, L., {Ekstr{\"o}m}, S., {Eggenberger}, P., {et~al.} 2024, \aap, 681,
  L1

\bibitem[{{Sciarini} {et~al.}(2026){Sciarini}, {Rosu}, {Ekstr{\"o}m},
  {Marchand}, {Eggenberger}, \& {Meynet}}]{sciarini26}
{Sciarini}, L., {Rosu}, S., {Ekstr{\"o}m}, S., {et~al.} 2026, arXiv e-prints,
  arXiv:2601.14363

\bibitem[{{Scott} {et~al.}(2021){Scott}, {Hirschi}, {Georgy}, {Arnett},
  {Meakin}, {Kaiser}, {Ekstr{\"o}m}, \& {Yusof}}]{scott21}
{Scott}, L.~J.~A., {Hirschi}, R., {Georgy}, C., {et~al.} 2021, \mnras, 503,
  4208

\bibitem[{{Scuflaire} {et~al.}(2008){Scuflaire}, {Th{\'e}ado}, {Montalb{\'a}n},
  {Miglio}, {Bourge}, {Godart}, {Thoul}, \& {Noels}}]{scuflaire08}
{Scuflaire}, R., {Th{\'e}ado}, S., {Montalb{\'a}n}, J., {et~al.} 2008, Ap\&SS,
  316, 83

\bibitem[{{Shakura}(1985)}]{shakura85}
{Shakura}, N.~I. 1985, Soviet Astronomy Letters, 11, 224

\bibitem[{{Song} {et~al.}(2013){Song}, {Maeder}, {Meynet}, {Huang},
  {Ekstr{\"o}m}, \& {Granada}}]{song13}
{Song}, H.~F., {Maeder}, A., {Meynet}, G., {et~al.} 2013, \aap, 556, A100

\bibitem[{{Song} {et~al.}(2016){Song}, {Meynet}, {Maeder}, {Ekstr{\"o}m}, \&
  {Eggenberger}}]{song16}
{Song}, H.~F., {Meynet}, G., {Maeder}, A., {Ekstr{\"o}m}, S., \& {Eggenberger},
  P. 2016, \aap, 585, A120

\bibitem[{{Spruit}(2002)}]{spruit02}
{Spruit}, H.~C. 2002, \aap, 381, 923

\bibitem[{{Sung} {et~al.}(1998){Sung}, {Bessell}, \& {Lee}}]{sung98}
{Sung}, H., {Bessell}, M.~S., \& {Lee}, S.-W. 1998, \aj, 115, 734

\bibitem[{{Sung} {et~al.}(2013){Sung}, {Sana}, \& {Bessell}}]{sung13}
{Sung}, H., {Sana}, H., \& {Bessell}, M.~S. 2013, \aj, 145, 37

\bibitem[{{Sweet}(1950)}]{sweet50}
{Sweet}, P.~A. 1950, \mnras, 110, 548

\bibitem[{{Tadross}(2003)}]{tadross03}
{Tadross}, A.~L. 2003, \na, 8, 737

\bibitem[{{Tkachenko} {et~al.}(2020){Tkachenko}, {Pavlovski}, {Johnston},
  {Pedersen}, {Michielsen}, {Bowman}, {Southworth}, {Tsymbal}, \&
  {Aerts}}]{tkachenko20}
{Tkachenko}, A., {Pavlovski}, K., {Johnston}, C., {et~al.} 2020, \aap, 637, A60

\bibitem[{{Vink} {et~al.}(2001){Vink}, {de Koter}, \& {Lamers}}]{vink01}
{Vink}, J.~S., {de Koter}, A., \& {Lamers}, H.~J.~G.~L.~M. 2001, \aap, 369, 574

\bibitem[{{{\v{S}}urlan} {et~al.}(2012){{\v{S}}urlan}, {Hamann}, {Kub{\'a}t},
  {Oskinova}, \& {Feldmeier}}]{surlan12}
{{\v{S}}urlan}, B., {Hamann}, W.-R., {Kub{\'a}t}, J., {Oskinova}, L.~M., \&
  {Feldmeier}, A. 2012, \aap, 541, A37

\bibitem[{{Zahn}(1977)}]{zahn77}
{Zahn}, J.-P. 1977, \aap, 57, 383

\bibitem[{{Zahn}(1992)}]{zahn92}
{Zahn}, J.~P. 1992, \aap, 265, 115

\end{thebibliography}


\begin{appendix} 
\section{Additional figures\label{appendix:figures}}

\begin{figure}[h!]
\centering
\includegraphics[width=\linewidth]{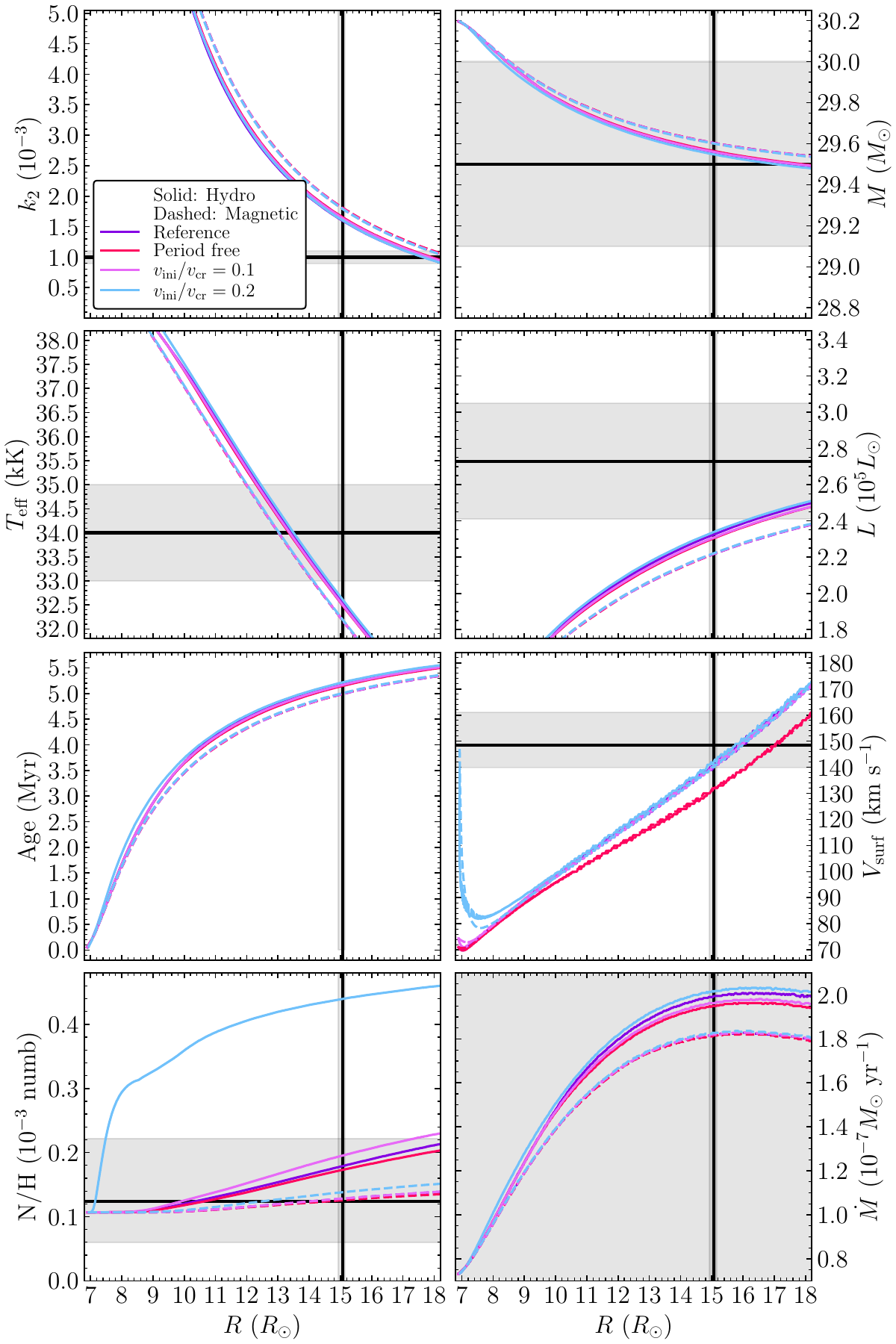}
\caption{Evolution of stellar parameters with stellar radius for hydro (plain lines) and magnetic (dashed lines) binary-star models for different values of $v_\text{ini}$ when not initialised synchronised, as well as models with orbital period not fixed. The models have $M_\text{ini} = 30.2M_\odot$, $\alpha_\text{ov} = 0.5$, $Z=0.0183$, $Y_\odot$, and \citet{krticka24} mass-loss rate. Models initialised synchronised, all other parameters identical, are shown for comparison. Observational values and their error bars are shown (plain black line and grey shaded area). \label{fig:notsynchro}}
\end{figure}

\begin{figure}
\centering
\includegraphics[width=\linewidth]{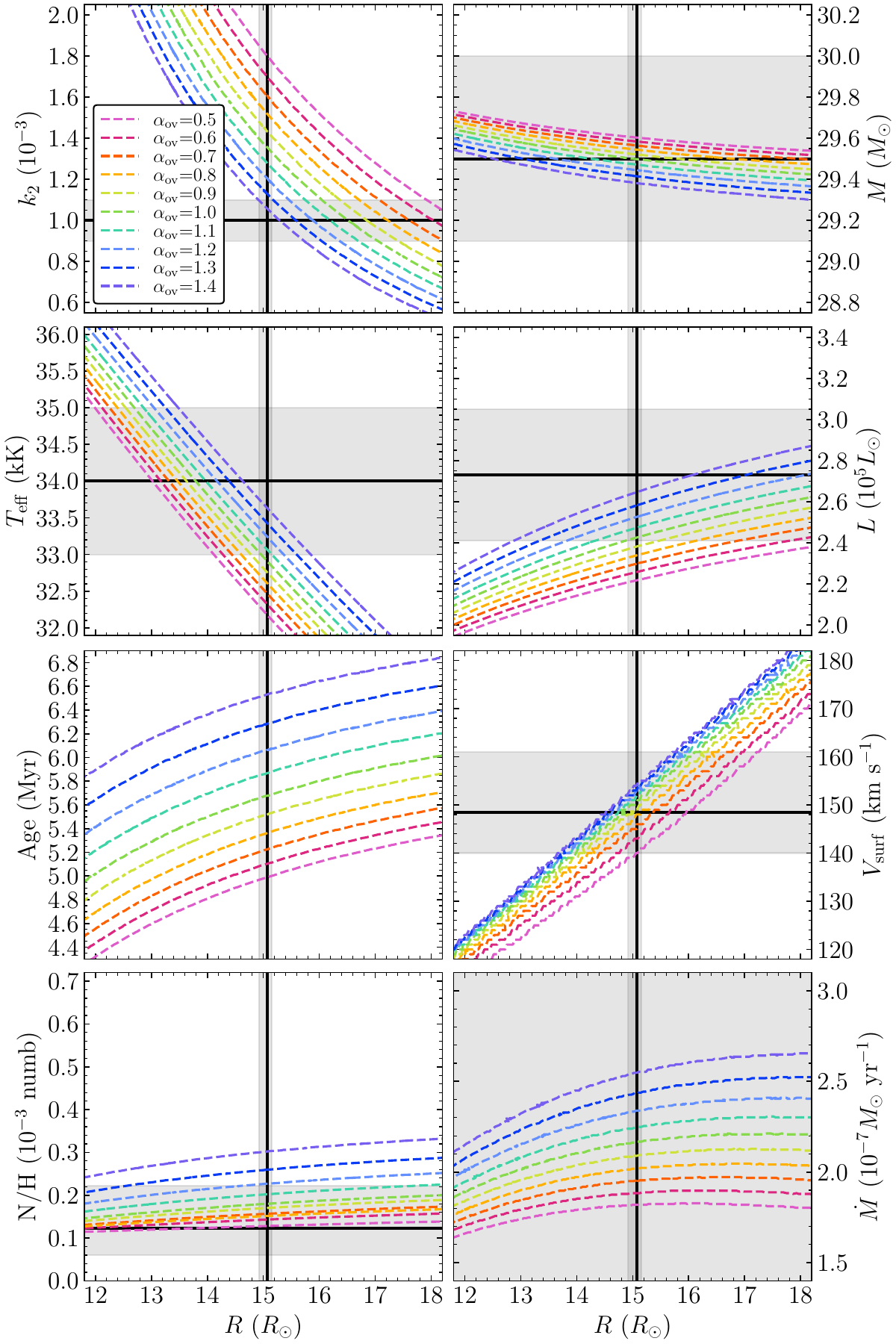}
\caption{Evolution of stellar parameters with stellar radius for magnetic binary-star models for different values of $\alpha_\text{ov}$. Models have $M_\text{ini} = 30.2M_\odot$, $Z=0.0183$, $Y_\odot$, and \citet{krticka24} mass-loss rate. Observational values and their error bars are shown (plain black line and grey shaded area). \label{fig:alpha_ov_mag}}
\end{figure}

\begin{figure}
\centering
\includegraphics[width=\linewidth]{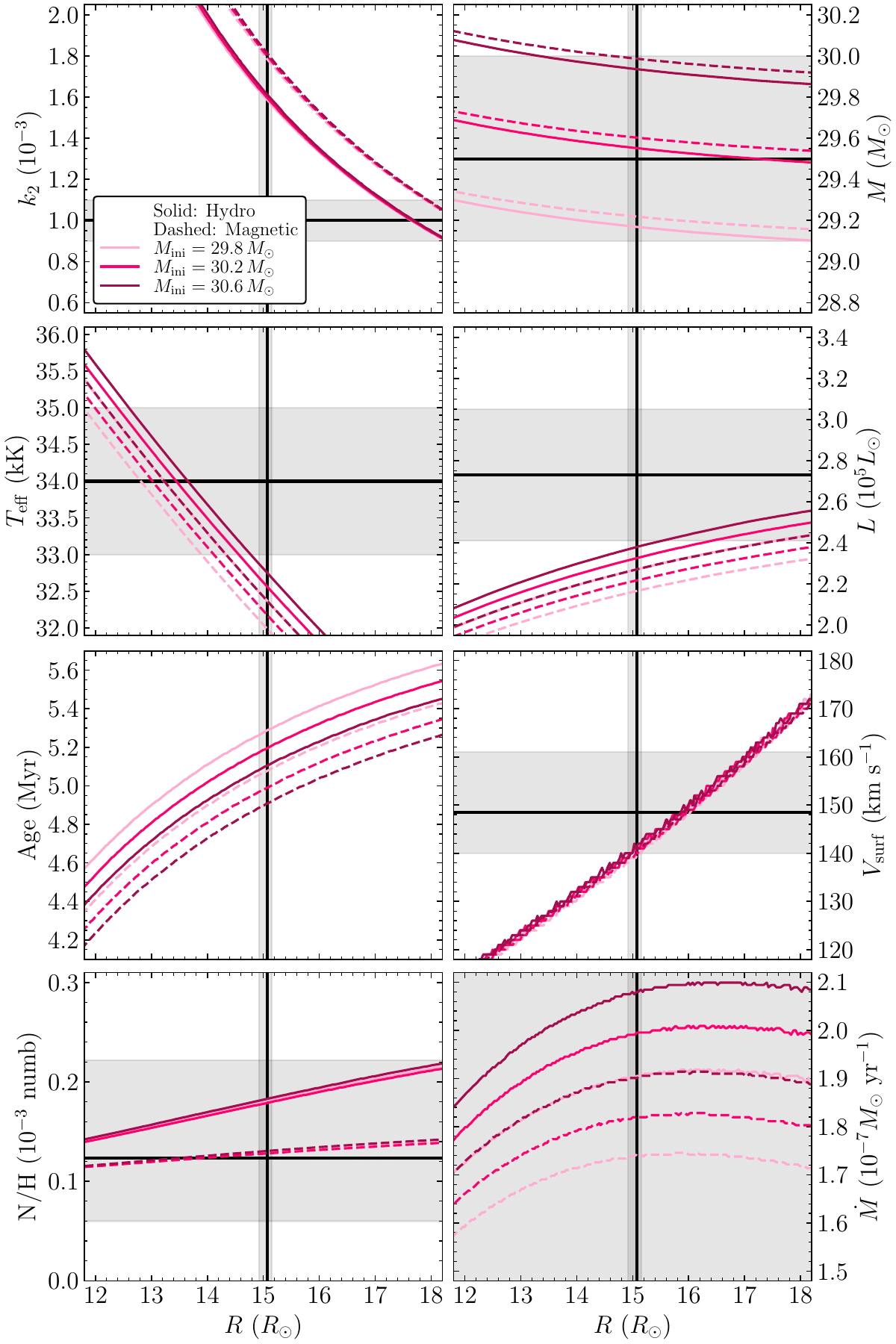}
\caption{Evolution of stellar parameters with stellar radius for hydro (plain lines) and magnetic (dashed lines) binary-star models for models with different initial masses. The models have $\alpha_\text{ov} = 0.5$, $Z=0.0183$, $Y_\odot$, and \citet{krticka24} mass-loss rate. Observational values and their error bars are shown (plain black line and grey shaded area). \label{fig:initial_mass}}
\end{figure}

\begin{figure}
\includegraphics[width=1\linewidth, clip=true, trim=0 0 0 0]{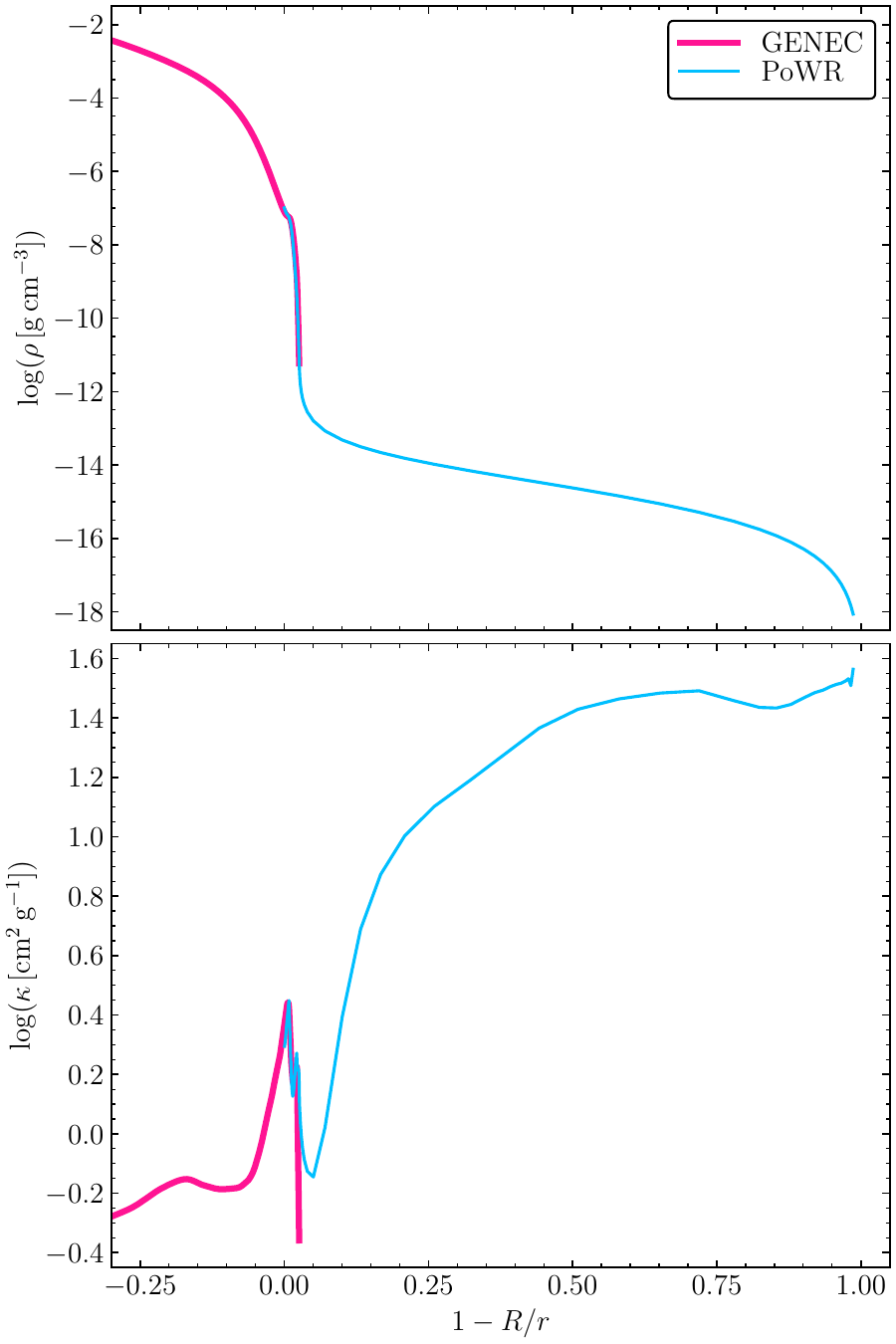}
\caption{Density and opacity profiles of a $M_\text{ini} = 30.2\,M_\odot$ star with $\alpha_\text{ov}=0.5$, $Z=0.0183$, $Y_\odot$, and \citet{krticka24} mass-loss rate of one model from \texttt{PoWR} and \texttt{GENEC}. The \texttt{GENEC} model uses a tabulated Rosseland-mean opacity, whereas the \texttt{PoWR} model computes the flux-weighted mean opacity explicitly given the radiation field and the electronic populations. \label{fig:powr_GENEC}}
\end{figure}

\begin{figure}
\includegraphics[width=\linewidth,, clip=true, trim=0 10 0 0]{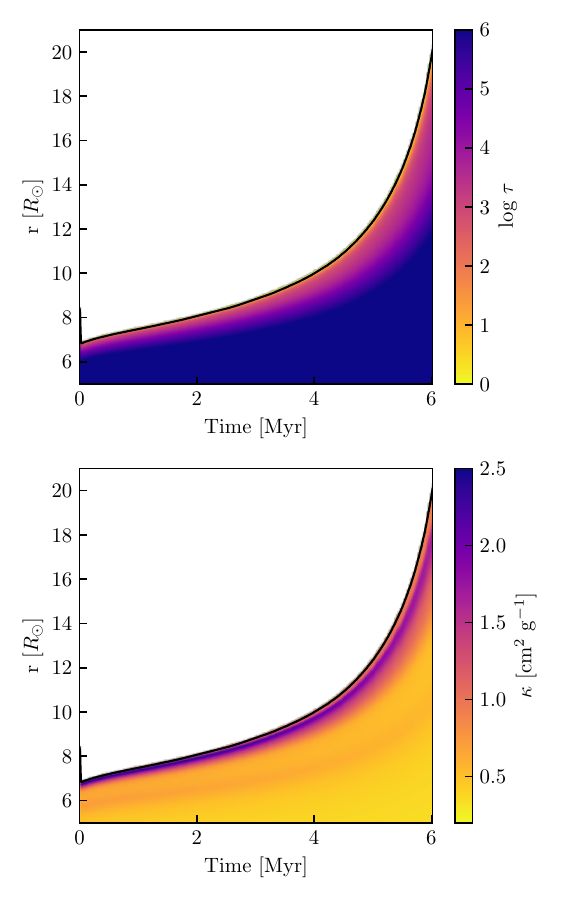}
\caption{Evolution of the interior atmospheric structure of the continuum optical depth (\textit{top panel}) and Rosseland mean opacity (\textit{bottom panel}) of a $M_\text{ini} = 30.2\,M_\odot$ star with $\alpha_\text{ov}=0.5$, $Z=0.0183$, $Y_\odot$, and \citet{krticka24} mass-loss rate during the main sequence.   \label{fig:powr}}
\end{figure}

\end{appendix}
\end{document}